\documentclass[Afour,sageh,times]{sagej}

\usepackage{moreverb,url}
\usepackage{colortbl}
\usepackage[utf8]{inputenc}
\usepackage{graphicx}
\usepackage{comment}
\usepackage[table]{xcolor}
\usepackage{multirow}

\usepackage[colorlinks,bookmarksopen,bookmarksnumbered,citecolor=red,urlcolor=red]{hyperref}

\newcommand\BibTeX{{\rmfamily B\kern-.05em \textsc{i\kern-.025em b}\kern-.08em
T\kern-.1667em\lower.7ex\hbox{E}\kern-.125emX}}

\def\volumeyear{2016}

\begin{document}

\runninghead{Ziemer and Schultheis}

\title{Three Orthogonal Dimensions for Psychoacoustic Sonification}

\author{Tim Ziemer\affilnum{1}\affilnum{2} and Holger Schultheis\affilnum{1}\affilnum{3}}

\affiliation{\affilnum{1}University of Bremen, Bremen Spatial Cognition Center \\
\affilnum{2}Medical Image Computing Group \affilnum{3}Institute for Artificial Intelligence}

\corrauth{Tim Ziemer, University of Bremen, Medical Image Computing Group, Enrique-Schmidt-Str. 5, 28359 Bremen, Germany}

\email{ziemer@uni-bremen.de}

\begin{abstract}
\textbf{Objective: } Three perceptually orthogonal auditory dimensions for multidimensional and multivariate data sonification are identified and experimentally validated. \textbf{Background:} Psychoacoustic investigations have shown that orthogonal acoustical parameters may interfere perceptually. The literature hardly offers any solutions to this problem, and previous auditory display approaches have failed to implement auditory dimensions that are perceived orthogonally by a user. In this study we demonstrate how a location in three-dimensional space can be sonified unambiguously by the implementation of perceptually orthogonal psychoacoustic attributes in monophonic playback. \textbf{Method:} Perceptually orthogonal auditory attributes are identified from literature research and experience in music and psychoacoustic research. We carried out an experiment with $21$ participants who identified sonified locations in two-dimensional space. \textbf{Results:} With just $5$ minutes of explanation and exploration, naive users can interpret our multidimensional sonification with high accuracy. \textbf{Conclusion:} We identified a set of perceptually orthogonal auditory dimensions suitable for three-dimensional data sonification. \textbf{Application:} Three-dimensional data sonification promises blind navigation, e.g. for unmanned vehicles, and reliable real-time monitoring of multivariate data, e.g., in the patient care sector.
\end{abstract}


\maketitle


\section{Introduction}
Sonification is a powerful means to complement or replace visual displays, especially in situations in which vision is limited (e.g., in darkness, fog, smoke, muddy waters, etc. or due to occlusion), in which the visual scene is overloaded (e.g. due to too many displays or visual distractors), or in which spatio-visual processing is the bottleneck of spatial cognition \cite{usepsycho}. 

There is a need for orthogonal dimensions in sonification for multidimensional or multivariate data \cite{stockmarket,9dimensions,barrassthesis,eyesfree,davidbuch} \cite[ch. 6]{davidworrall}. The most prominent application area for multidimensional sonification is spatial navigation, e.g., in real and virtual environments \cite{jmui,matti,vadis}, games \cite{walkinggame}, piloting \cite{vdisflight,blindflight}, remote vehicle control \cite{remote}, autonomous driving\cite{colli}, image-guided surgical interventions \cite{davidstandalone,asa,poma} and neuronavigation \cite{actaneurochir}. \cite{degara} even consider sonification for navigation ``one of the most important tasks in auditory display research''. Besides navigation, examples for multidimensional or multivariate data sonification include motion analysis and interactive feedback in sports training and neuromotor rehabilitation \cite{reha,stroke,hip,rehaplan} 
 and in multivariate data monitoring, like anesthesia and patient monitoring \cite{patientmonitoring}, stock market monitoring \cite{stockmarket}, data exploration and browsing \cite{9dimensions,datamining,osti_5221536,rebeccastewart,dimensions,thomas}. 

Here, perceptual orthogonality means that while two quantities are simultaneously sonified, both can be interpreted. Moreover, if one quantity changes, the change of sound can be attributed to its corresponding quantity, and unambiguously interpreted. This obvious necessity is not easily achieved. Due to the complicated, nonlinear processing of the auditory system, all physical sound field quantities can affect practically all perceptual attributes of sound. Despite its importance, the lack of perceptual orthogonality is considered one of the most challenging issues in \emph{sonic interaction design}, 
\emph{auditory interfaces for Human-Computer Interaction}, \emph{auditory display}, and, especially, \emph{sonification} design \cite{sonicinteraction,hcihb,davidworrall,thomas,123channels2,123channels,neuortho,kramer,grond}.

In this paper we present auditory attributes that can serve as three orthogonal dimensions. The approach is evaluated in a listening test with naive listeners.

\section{Background}
Orthogonality is a topic that has been treated a lot in the fields of psychoacoustics and auditory display research, and will be briefly discussed in this section, followed by previous work. A lot of previous work either focused on the implementation of psychoacoustics in sonification design or on orthogonal dimensions in sonification. Our work integrates these two lines of research by leveraging psychoacoustic knowledge to sonify perceptually orthogonal dimensions. 
\subsection{Orthogonality}
Following the literature on sonification \cite{davidworrall,thomas} and psychoacoustics \cite{schneid1,schneiderfunda} we define dimensions in a Cartesian way as having both a \emph{direction} and a \emph{distance}, also referred to as \emph{polarity} and \emph{magnitude}. In that sense, information like the radius, which is one dimension in polar and cylindrical coordinates, is only a \emph{half-dimension}; it only informs about a distance, not a direction \cite{guidelinespsy}. It is widely accepted that auditory sensations and other psychological attributes are never perfectly orthogonal, as the dimensions may be correlated to some extent \cite{schneid1}. Hence, we consider dimensions as orthogonal, if a magnitude change of one dimension hardly affects the magnitude of any other dimension, often referred to as \emph{separable} \cite{garner,schneid,ecol}. In that sense, they are linearly independent from one another \cite{davidworrall}, i.e., they barely exhibit any \emph{coupling} or \emph{perceptual interactions} \cite[ch. 3]{thomas}; \cite{123channels,123channels2}. Furthermore, a dimension must be continuous, i.e., on interval scale or ratio scale rather than ordinal or nominal scale \cite{schneid1}.
\subsection{Orthogonality in Sonification}
There are plenty examples of sonifications mapping one half-dimension to amplitude and another to frequency \cite{stockmarket} 
and it is not surprising that the authors realized in their evaluations that these physically orthogonal dimensions interact perceptually.

\cite{123channels2} tried out several mapping  principles for multivariate data in complex work domains. They do not consider psychoacoustics in their parameter mapping approach, but map multiple variables to physical parameters, like amplitude, amplitude modulations, fundamental frequency, cutoff-frequency, pulse width, etc. They realize that participants had problems interpreting multiple variables at once. They criticize that psychoacoustic research does not provide sufficient guidelines for sophisticated, orthogonal sonification design. Yet, they hope that ``...careful sonification design, based on a complete understanding of the mechanisms causing perceptual interactions, could overcome such problems''. A similar observation has been made by \cite{warningdesign}, who carried out experiments in which they altered the magnitude of several acoustical quantities to see how it affects perceived urgency. They realized that changing the magnitude of one parameter, like raising the fundamental frequency, increasing the amplitude or increasing the playback speed, increased the perceived urgency. However, when altering several parameters at once, the urgency levels do not add up, but create somewhat nonlinear effects. 

\cite{thomas} lists the ``lack of perceptual orthogonality'' as one of the most important difficulties in auditory displays. He agrees with \cite{123channels2} that auditory perception is too little understood to specify multiple orthogonal dimensions. Likewise, \cite{hcihb} lists the ``lack of orthogonality'' 
one of the problems with nonspeech sound.

\cite[ch. 2]{davidworrall} agrees with this observation, too, stating that ``(\ldots) parameter mapping requires a working knowledge of how the parameters interact with each other perceptually'', because these interaction may obscure data relations and even confuse the listener \cite{davidbuch}. He thinks that attempts to create a perceptually orthogonal sonification space have not yet been successful, giving the timbre space sonification approach \cite{barrassthesis} as an example. At the same time, he expressed the need for better tools.



\subsection{Psychoacoustics in Sonification Design}
The need to consider psychoacoustics in sonification design has been expressed in numerous studies \cite{warningdesign,thomas,sonifilab,osti_5221536,kramer,framework,smith1990,doasa,evaluation,walkerkramerpsy,psysoni,trustus,arbitrary,nagelsonex}.

\cite{evaluation} evaluate psychoacoustic parameters for sonification. They argue that pitch is a meaningful dimension, as human listeners have a high resolution in pitch perception. They suggest the use of loudness fluctuation and roughness as additional dimensions. 
The authors of \cite{psysoni} suggest mapping of parameters to pitch, loudness, roughness, and brightness. Likewise, \cite{guidelinespsy} consider pitch, loudness, duration/tempo and timbre as orthogonal and as the main perceptual attributes of sound. \cite{musicmap} name pitch, loudness, timbre aspects, like brightness, roughness, attack time, vibrato and formants, spatialization, as well as their temporal derivatives, as psychoacoustic parameters suitable for understandable multidimensional sonification.


The authors of \cite{guidelinespsy} managed to implement and validate sonification designs derived from psychoacoustic considerations. Here, the distance to a target was not just mapped to physical audio parameters, but to psychoacoustic quantities. The distance to a target was mapped to the speed of modulations of either frequency or amplitude. These modulations create the impression of pitch fluctuations or loudness fluctuations, respectively. Only at the target location, the pitch, or loudness, respectively, was steady. They implemented neither a complete one-dimensional approach (with both a polarity and a distance) nor a multi-dimensional approach. But they suggest to map orthogonal dimensions to segregate auditory streams \cite{bregman}, like one to pitch- and another to tempo-fluctuations.
\subsection{Orthogonal Psychoacoustic Sonification}
A few studies aimed at creating multi-dimensional sonification based on psychoacoustic knowledge. Already in 1980 \cite{9dimensions} argues that no less than $36$ dimensions can be created from the parameters pitch, loudness, damping, direction, duration of sound and duration of silence, attack time, phase coherence and overtones. 
 However, the study gives no evidence for this claim. The work does not clearly distinguish between physical and perceptual parameters and neglects the interference problems mentioned in the previous section.

\cite{framework,barrassthesis} describes a theory to map three cylindrical dimensions to a perceptual auditory space. Here, pitch is the height dimension, brightness is the radius, and timbre in terms of different musical instruments serve as angles. However, the author identify timbre in terms of musical instruments to be nominal rather than in interval scale. Furthermore, he recognized that this timbre choice does not allow for comprehensible opposite angles, which would be necessary for interpretable cylindrical coordinates. Overall, he considered the sounds of his sonification approach as difficult to interpret.



\cite{stroke} map one direction to pitch and another one to brightness of a synthesized sound. The sonification informs about the magnitude in each direction, but not about the polarity. Hence, we consider these as half-dimensions. In an experiment elderly participants were presented one reference sound. Then, they explored a map with $7$ times $7$ fields, each playing one sound with a distinct combination of pitch and brightness. Their task was to select the field whose sound equaled the reference sound. Their mean error lay between about $0.3$ and $0.7$ fields for the pitch direction and between $1$ and $1.6$ for the brightness direction. A random guess would have led to a mean error of $2.2$. Based on these results, they consider the two parameters as orthogonal and implement the two, together with loudness as parameter for the third half-dimension, for motion sonification in \cite{10.3389/fneur.2016.00106}. However, their study does not evaluate the perceptual orthogonality of the third dimension.



The authors of \cite{psysoni} come up with a framework, for psychoacoustic sonification of multidimensional or multivariate data. They suggest to map one dimension or variable to one psychoacoustic parameter and another dimension or variable to another psychoacoustic parameter. They understand that mapping orthogonal data to the magnitude of orthogonal auditory qualities is an inverse problem; the desired perceptual outcome is known, but the the physical audio parameters necessary to create such output need to be found. This problem is ill-posed. Hence, there is no analytical solution. They suggest to solve the problem by massive lookup tables. However, they see the problem that this solution may cause  large jumps of audio parameter magnitudes by just small changes of the input data. These jumps may cause audible artifacts.


In our own previous work we introduced chroma as one and a combination of beats and roughness as orthogonal auditory dimensions for two-dimensions sonification \cite{cars,dgm}. The digital signal processing for this psychoacoustic sonification approach is explained in \cite{poma,acta}. We validated the approach in a passive listening experiment with $7$ inexperienced listeners. In a multiple-choice task with $16$ fields on a map, they correctly identified $41$\% of the sonified targets even though the performance of one participant was near chance level. Over $83$\% of the quadrants were identified correctly. These results indicate that listeners are able to interpret the direction and distance along each dimension independently, despite the fact that both dimensions are presented at the same time. Motivated by these results we carried out slight improvements of the sonification and then conducted an interactive experiment with $18$ participants \cite{ziemerschicad,jmui}. Results of this experiment underlined that these dimensions are in fact orthogonal, and gave additional indication about good linearity and high resolution of the dimensions and about learnability and training effects and the way people interact with the sound in a navigation task. In \cite{icad2019} we describe a modified signal processing approach to add a third dimension to the two-dimensional sonification. However, the interpretability and orthogonality have only been explored by the authors.

In this contribution we explain how we derived the third dimension. We repeat the passive experiment from \cite{poma,cars} to evaluate the orthogonality of our improved two-dimensional sonification and our newly introduced third dimension as described in \cite{icad2019}.
\section{Psychoacoustic Sonification}
This section starts with an overview of perceptual auditory qualities that can be found in the literature. We then describe how to derive three orthogonal dimensions, including direction and distance. We distinguish acoustic attributes from auditory attributes, the first describing the physical domain, the latter referring to the perceptual domain.

\subsection{Perceptual Auditory Qualities}
Previous work demonstrated that acoustic attributes may interfere perceptually, and even individual auditory qualities may correlate to some extent. This led to the above statements that there is a lack of orthogonal auditory attributes\cite{thomas,davidworrall,123channels2}. However, in our opinion, orthogonal auditory dimensions exist. What is missing is a comprehensive treatise of orthogonality in the psychoacoustic literature. As \cite{ecol} states: ``\ldots perceptual interaction of auditory dimensions (\ldots) have also been studied very little compared with more traditional areas of psychoacoustic research''. However, a heuristic technique to derive perceptually independent attributes for multidimensional sonification is an accepted and promising approach \cite{davidbuch}. A brief discussion is presented in this section.

Literature on auditory sensation and perception describes several auditory qualities. Some are unidimensional, others are multidimensional. Some are independent from the others, whereas some interfere to some extent. Auditory qualities include:

\begin{itemize}
\setlength\itemsep{-0.2em}
\item Loudness \cite[ch. 8]{zwicker}; \cite[ch. 4]{ziemer}
\item Pitch \cite[ch. 5]{zwicker}; \cite{shepard}; \cite[ch. 4]{ziemer}; \cite{ecol}
\begin{itemize}
\item[$\ast$] Height \cite{shepard}; \cite[ch. 4]{ziemer}; \cite{schneiderpitch}
\item[$\ast$] Chroma \cite{shepard}; \cite[ch. 4]{ziemer}; \cite{schneiderpitch}
\item[$\ast$] Strength/Salience \cite[ch. 5]{zwicker}; \cite[ch. 4]{ziemer}; \cite{schneiderpitch}
\end{itemize}
\item Timbre \cite{schneid}; \cite[ch. 4]{ziemer}; \cite{ecol}
\begin{itemize}
\item[$\ast$] Sound color/tonal color \cite{schneid}; \cite[ch. 4]{ziemer}
\item[$\ast$] Brightness/sharpness \cite{zwicker,schneid}
\item[$\ast$] Roughness/sensory dissonance \cite[ch. 11]{zwicker}; \cite{schneid}
\item[$\ast$] Percussiveness vs. mellowness \cite{schneid}
\item[$\ast$] Tonalness/tonality vs. noisiness \cite[ch. 9]{zwicker}; \cite{schneid}
\item[$\ast$] Harmonicity \cite[ch. 4]{ziemer}; \cite{schneid}
\item[$\ast$] Fullness/volume/sonority \cite{schneid}
\item[$\ast$] Beats / Loudness fluctuation \cite[ch. 4]{ziemer}; \cite[ch. 4]{zwicker}
\item[$\ast$] Vowel quality/Vocality \cite{schneid,schneiderpitch}
\end{itemize}
\item Sensory pleasantness \cite[ch. 9]{zwicker}
\item Auditory event location \cite[ch. 15]{zwicker}; \cite[ch. 4]{ziemer}; \cite{ecol}
\begin{itemize}
\item[$\ast$] Azimuth angle \cite[ch. 4]{ziemer}
\item[$\ast$] Median angle \cite[ch. 4]{ziemer}
\item[$\ast$] Distance \cite[ch. 4]{ziemer}
\end{itemize}
\item Perceived source extent \cite[ch. 4]{ziemer}
\item Rhythm \cite[ch. 13]{zwicker}
\begin{itemize}
\item[$\ast$] Subjective duration \cite[ch. 12]{zwicker}; \cite{schneid}
\item[$\ast$] Fluctuation strength \cite[ch. 10]{zwicker}
\end{itemize}
\end{itemize}

In the psychoacoustic literature orthogonality of only a few of the above-listed auditory attribute has been discussed, e.g., in \cite{fire,aureswohl,orthogopitch,schneid,schneid1,schneiderpitch,shepard,grau,ecol,zwicker,terh,lichte}.


From our own experience in the recording studio and in psychoacoustic research, and from the above-mentioned literature, we could already combine roughness and the subjective duration of chroma change and beats to two dimensions and provide evidence for their orthogonality \cite{poma,dgm,cars}. These two dimensions are briefly described in Sect. \emph{Orthogonal Qualities}.

To extend our previous sonification to three dimensions, the literature suggest the use of sharpness, tonalness, and/or fullness. Unfortunately, tonalness is not an option, since a low degree of tonalness, i.e., a high degree of noisiness eliminates pitch in terms of both height and chroma. This means that tonalness is not orthogonal to any aspect of pitch. This leaves us mainly one choice: the incorporation of sharpness and fullness for the two directions of the third dimension. The signal processing to implement this has been described in \cite{icad2019}.

The paper at hand explains how the sonification works and gives an experimental evaluation of the orthogonality of these perceptual auditory attributes. Experimental results provide evidence that all dimensions are independent from one another. The sonification is perceived as one sound, i.e., as one auditory stream in terms of auditory scene analysis \cite{asa}. This sound has multiple orthogonal characteristics, i.e., independent perceptual auditory qualities. Each perceptual auditory quality represents another direction along the orthogonal dimensions within the Cartesian space. The magnitude of each individual quality indicates the distance along that direction.

\subsection{Orthogonal Qualities}
\label{ortho}
From the discussion above, we can derive an orthogonal, three-dimensional sonification. Following the idea of \cite{psysoni,hciindi}, we map three orthogonal dimensions to independent psychoacoustic parameters, which are closely related to perceptual auditory qualities. Each quality stands for one direction, its magnitude for the distance along that direction. As most perceptual auditory qualities tend to have a magnitude, but no direction, two of the three dimensions are made of two independent perceptual auditory qualities, one for each direction along the dimension. The result is a three-dimensional sonification made from five psychoacoustic quantities that are summarized in table \ref{tab:xyz}.

\begin{table}[!h]
\begin{tabular}{|c|c|c|c|}
\hline 
Dim. & Dir. & Psychoac. quantity & Dist.\\ 
\hline 
\multirow{2}{*}{$x$} & left & counterclockwise chroma change & speed \\ 
& right & clockwise chroma change & speed \\ 
\hline 
\multirow{2}{*}{$y$}  & up & loudness fluctuation & speed \\ 
& down & roughness & degree \\ 
\hline 
\multirow{2}{*}{$z$} & front & fullness & degree \\ 
& back & brightness & degree  \\ 
\hline 
\end{tabular} 
\caption{Psychoacoustic mapping for orthogonal multidimensional data sonification, including dimension (Dim.), direction (Dir.) and distance (Dis.).}
\label{tab:xyz}
\end{table}

\begin{figure}[!h]
\centering
\includegraphics[width=0.45\textwidth]{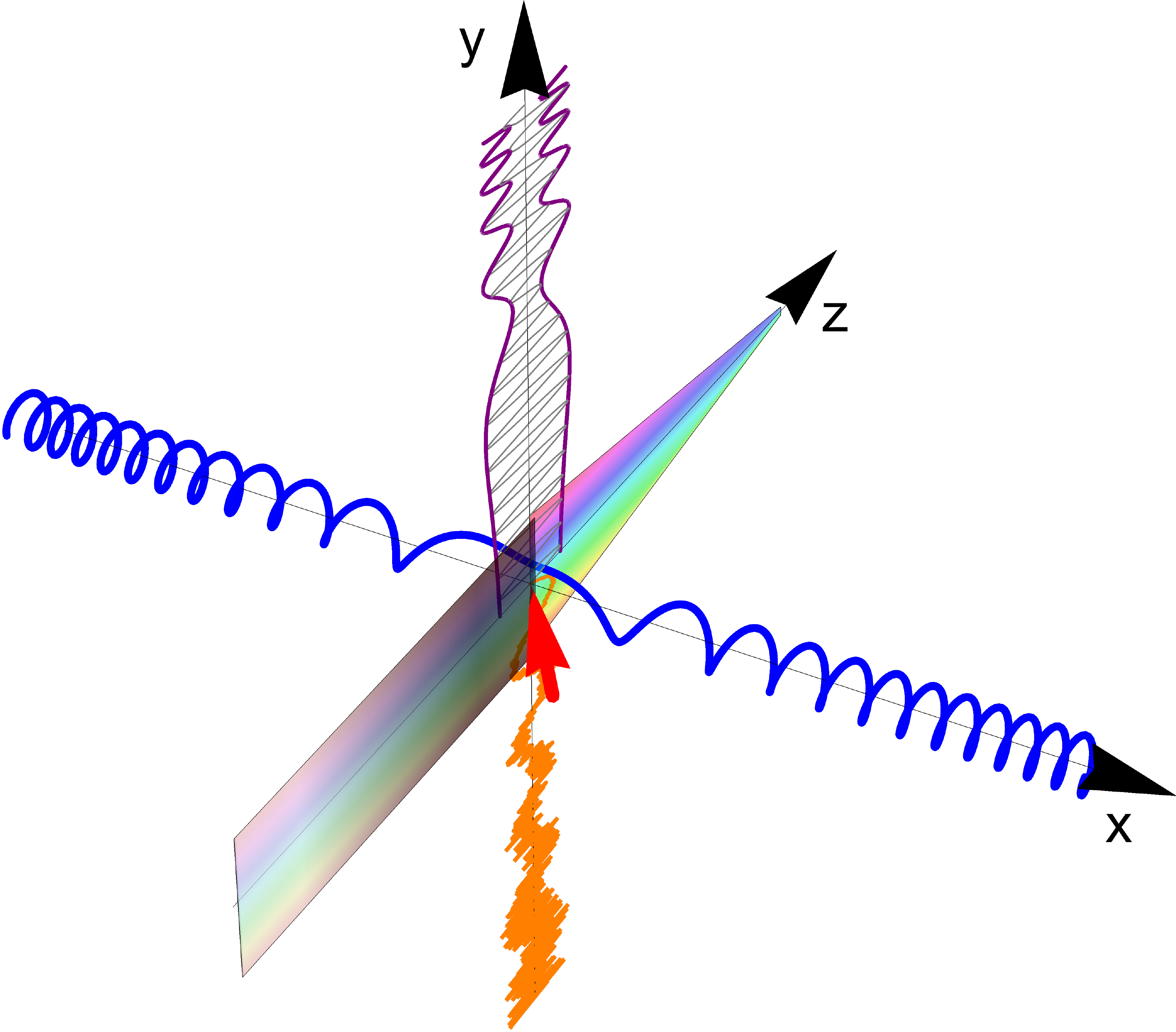}
\caption{Psychoacoustic sonification principle. The $x$-dimension is related to chroma, the $y$-dimension to beats and roughness and the $z$-dimension to fullness and brightness. In this illustration, the origin of the coordinate system is the location of the user (red arrow).}
\label{pic:iwin}
\end{figure}

The detailed signal processing for the psychoacoustic sonification is described in \cite{acta,icad2019}. Figure \ref{pic:iwin} illustrates the three-dimensional sonification. The $x$-axis is the chroma axis. At $x=0$, the pitch is steady in terms of both chroma and height. Targets to the right are denoted by a clockwise motion of chroma. Most listeners perceive this as a rising pitch \cite{shepard}. The further to the right, the faster the chroma cycles clockwise. In the figure the rising speed of clockwise chroma change is indicated by the blue, clockwise winding whose density of turns increases. Targets to the left are denoted by a counterclockwise motion of chroma. Most listeners perceive this as a falling pitch \cite{shepard}. The further to the left, the faster the chroma cycles counterclockwise. In the figure the counterclockwise chroma change is indicated by the blue, counterclockwise winding. 
The $y$-axis is divided in two. A target above is indicated by cyclic, continuous loudness fluctuation. The distance is denoted by the fluctuation speed. The further up, the faster the fluctuation. In the graphic this is indicated by the purple envelope with increasing fluctuation density. A target below is denoted by roughness. The further down, the higher the degree of roughness. In the graphic this is indicated by the orange curve that fades from sinusoidal to random. The $z$-dimension is also divided in two. Targets in front are denoted by fullness. The further away, the lower the degree of fullness. In the graphic this is indicated by the rainbow whose spectral bandwidth decreases. Targets in the back are denoted by brightness. The distance in this direction is denoted by the degree of brightness. In the graphic this is indicated by the visual brightness level of the rainbow.

Note that the sonification is perceived as one continuous sound, i.e., as one \emph{auditory stream} in terms of auditory scene analysis \cite[ch. 4]{ziemer}; \cite{asa}. 
The magnitude of its perceptual auditory qualities inform about the distance along its respective direction. No reference sound is needed, as the sonification itself communicates if the target is already reached, and if not, where it is located.

\section{Evaluation}
In this section we describe our experimental setup to evaluate the orthogonality of the proposed dimensions for psychoacoustic sonification. Basically, we employed the same experimental setup as in of our previous study with passive listeners \cite{poma,cars,dgm}. These existing results serve as a benchmark. 

We repeated the experiment for the $x$-$y$-plane to ensure that the modified signal processing did not affect the interpretability of these two dimensions. More importantly, we carried out the same experiment for the $x$-$z$ and the $z$-$y$ plane to evaluate whether the new $z$-dimension is readily interpretable and orthogonal to both the $x$- and the $y$-dimension. We decided to stick to two dimensions at a time because this procedure is typical for evaluating orthogonality of auditory attributes \cite{ecol}, since learning two attributes is easier for inexperienced listeners than learning three attributes. Furthermore, we already have a benchmark for two- but not for three-dimensional sonification.

We carried out the experiment with $N=21$ participants ($4$ female, age between $20$ and $53$, median = 26, mean = $27.8$, $\sigma=8.5$). Most participants were recruited from our near environment, i.e., mostly undergraduate and graduate computer science students. Participants volunteered to take part in the study without monetary compensation. First, the participants filled out a questionnaire, reporting age and sex, confirming that they were not aware of suffering from hearing loss, and rating their previous experience with sonification on a scale from $0$ (no experience) to $6$ (a lot of experience).
Some of the participants had heard previous versions of the sonification, or were familiar with sonification, generally,  from their car's park distance control system, whereas others were completely naive concerning sonification. The participants were arbitrarily assigned to one of the three groups $x$-$y$, $x$-$z$ and $z$-$y$, so that each group comprised $7$ participants.

To each group we first explained the psychoacoustic mapping principle, which took about $5$ minutes. First, we explained the sound attributes for the horizontal dimension in colloquial terms and imitated it with our voice. We repeated this for the vertical dimension. Then, in contrast to our earlier study, we let the participants explore the two dimensions themselves with a computer mouse for about $2$ minutes. Our hope was that this interaction with the sound would create a better understanding of the sonification, so no participant would perform at chance level.

\begin{figure}
\centering
\includegraphics[width=0.45\textwidth]{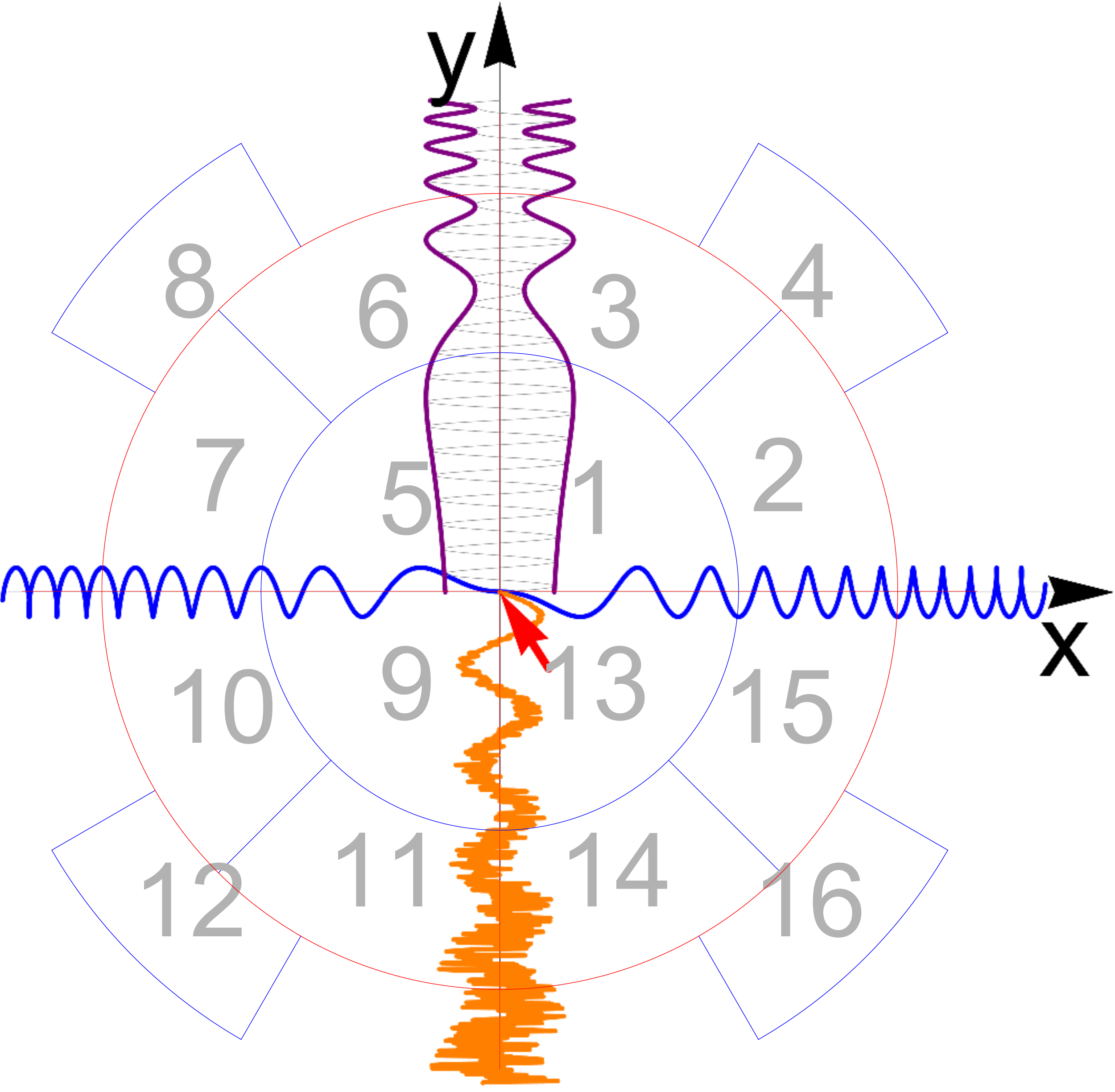}
\caption{Map and psychoacoustic sonification metaphors for the $x$-$y$ group. The target fields are numbered from $1$ to $16$.}
\label{pic:xy}
\end{figure}

\begin{figure}
\centering
\includegraphics[width=0.45\textwidth]{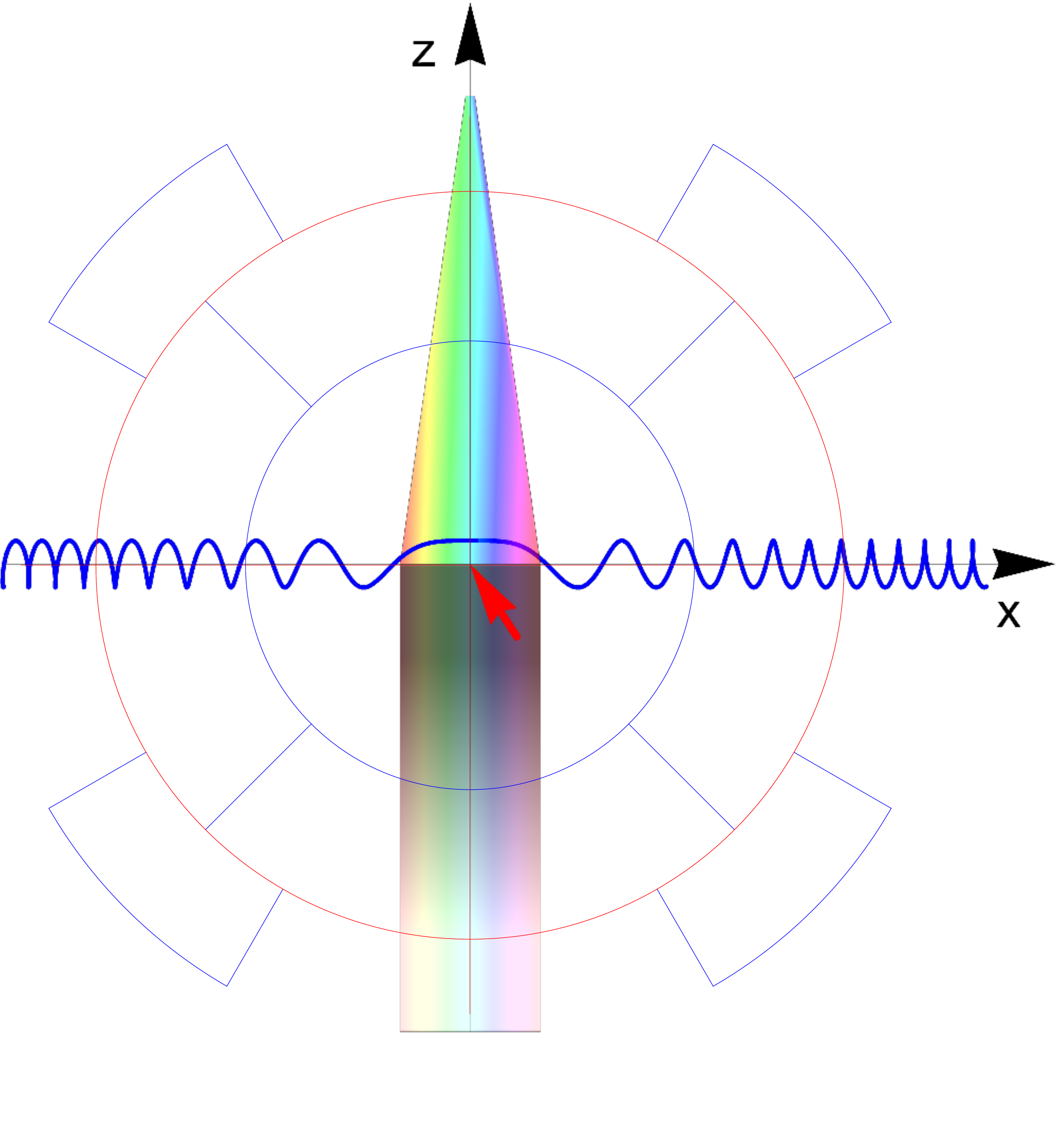}
\caption{Map and psychoacoustic sonification metaphors for the $x$-$z$ group.}
\label{pic:xz}
\end{figure}

\begin{figure}
\centering
\includegraphics[width=0.45\textwidth]{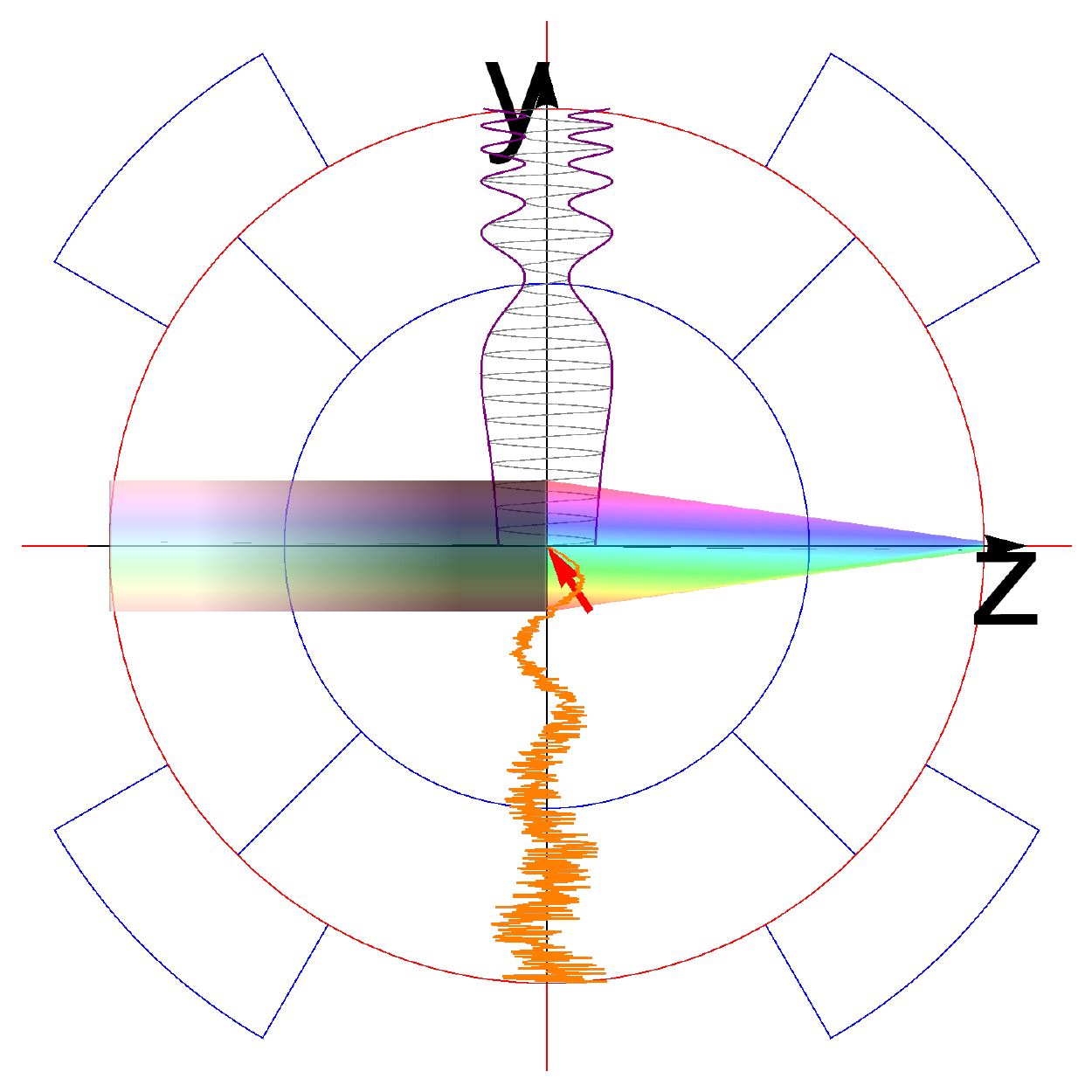}
\caption{Map and psychoacoustic sonification metaphors for the $z$-$y$ group.}
\label{pic:zy}
\end{figure}

We explained the experimental procedure to the participants. We showed them a map with $16$ fields as illustrated in the background of Figs. \ref{pic:xy} to \ref{pic:zy}. Then, a series of $20$ sounds was played to them. Each sound was a sonification of a location in one of the fields. They could take all the time they needed to decide in what field they assumed the sonified target to be and click on it. After each click, the next target was sonified without a pause in between. Participants did not receive feedback on their choice. We told them that the order of sonified targets would be random and that a) one or more target fields might be sonified multiple times and b) not necessarily every target field would be sonified. We did this to prevent the participants from drawing conclusions from already experienced trials, like excluding fields that they had clicked before. In fact we sonified all $16$ target fields in pseudo-random order and then repeated four randomly chosen fields. Participants were allowed to adjust the volume as they like and even mute the sound occasionally, if it would help them to take a break or to concentrate better.

The experimental preparation, i.e., explanation of the mapping principle, the sonification exploration and the description of the experiment process took about $8$ minutes.

\section{Results and Discussion}
\begin{figure}[thpb]
\centering
\includegraphics[width=0.45\textwidth]{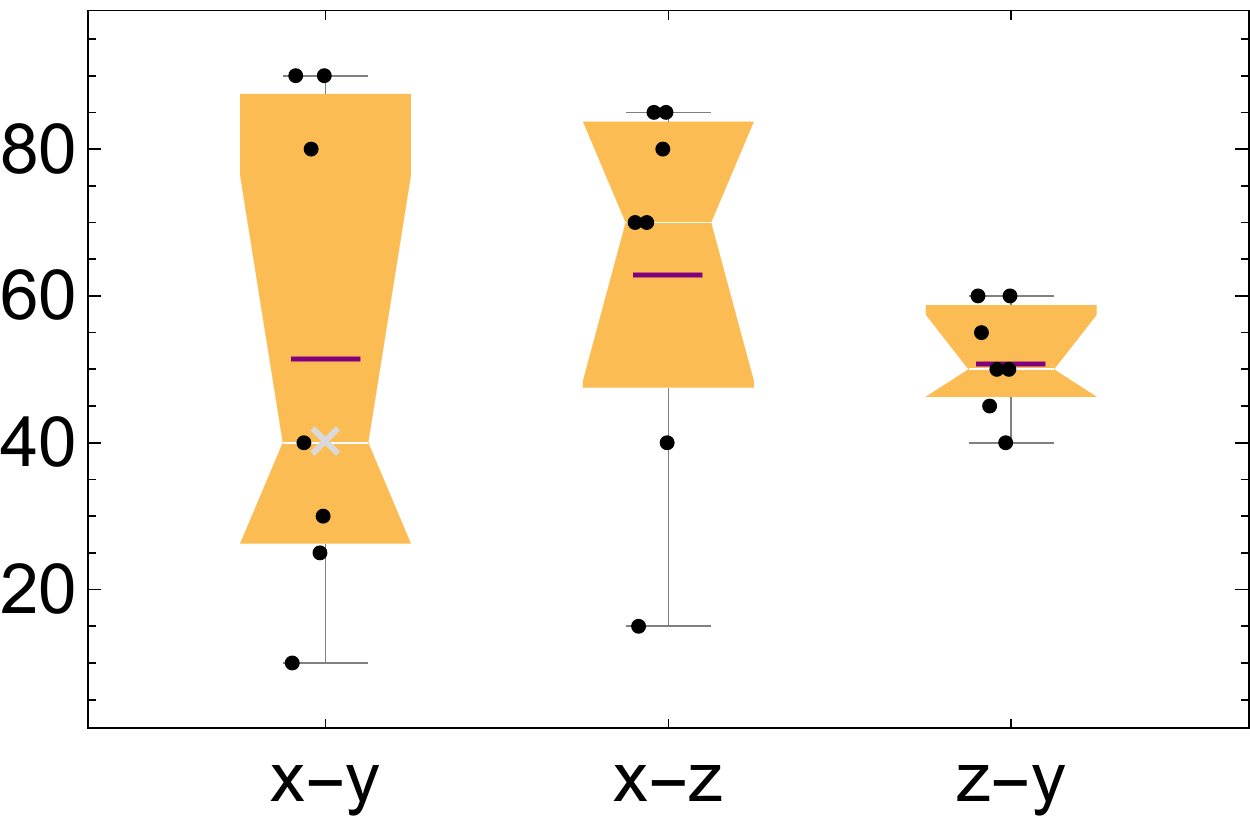}
\caption{Boxplot of hits per group, showing minimum and maximum (whiskers), median (white dash) and arithmetic mean (dark dash), the performance of each individual participant (dots), and the arithmetic mean (gray x) from our benchmark study \cite{poma,cars,dgm}.}
\label{pic:boxhit}
\end{figure}

\begin{figure}[thpb]
\centering
\includegraphics[width=0.45\textwidth]{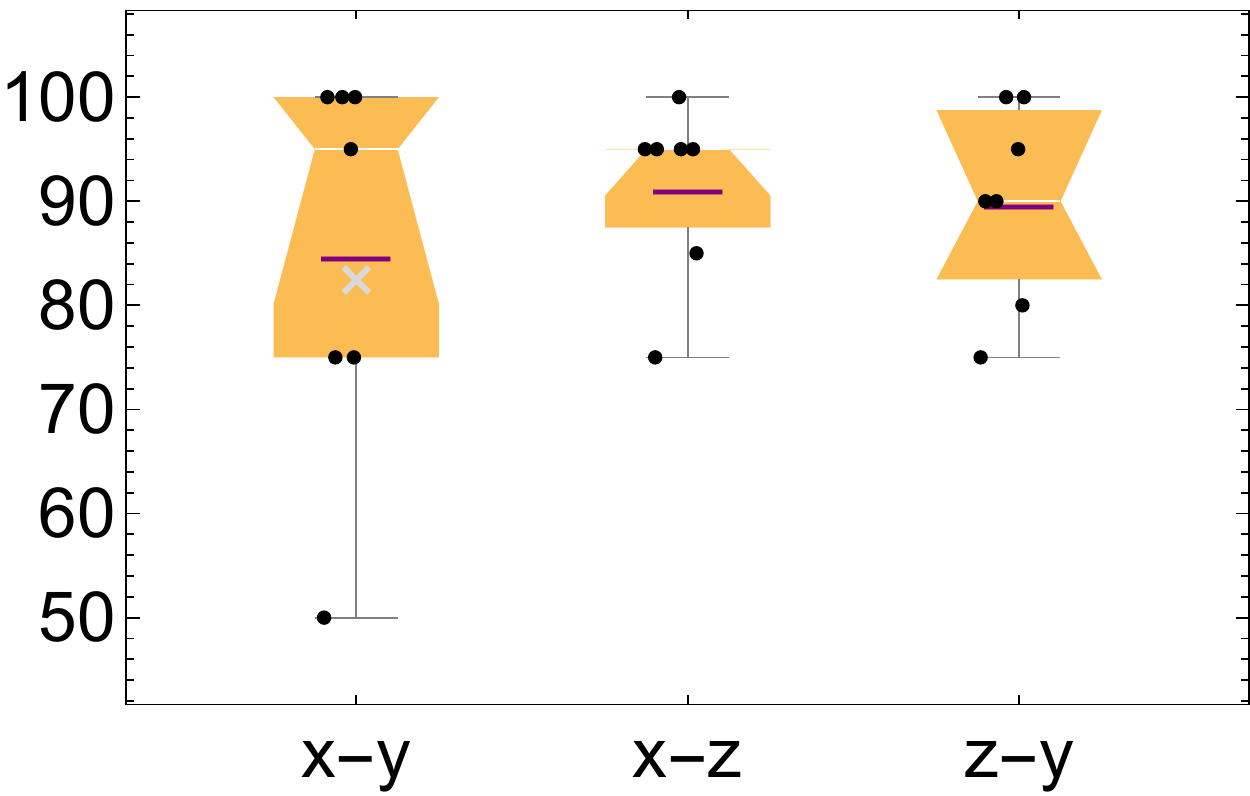}
\caption{Boxplot of correctly identified quadrants per group, showing minimum and maximum (whiskers), median (white dash) and arithmetic mean (dark dash), the performance of each individual participant (dots) and the arithmetic mean (gray x) from our benchmark study \cite{poma,cars,dgm}.}
\label{pic:boxq}
\end{figure}

\begin{figure}[thpb]
\centering
\includegraphics[width=0.45\textwidth]{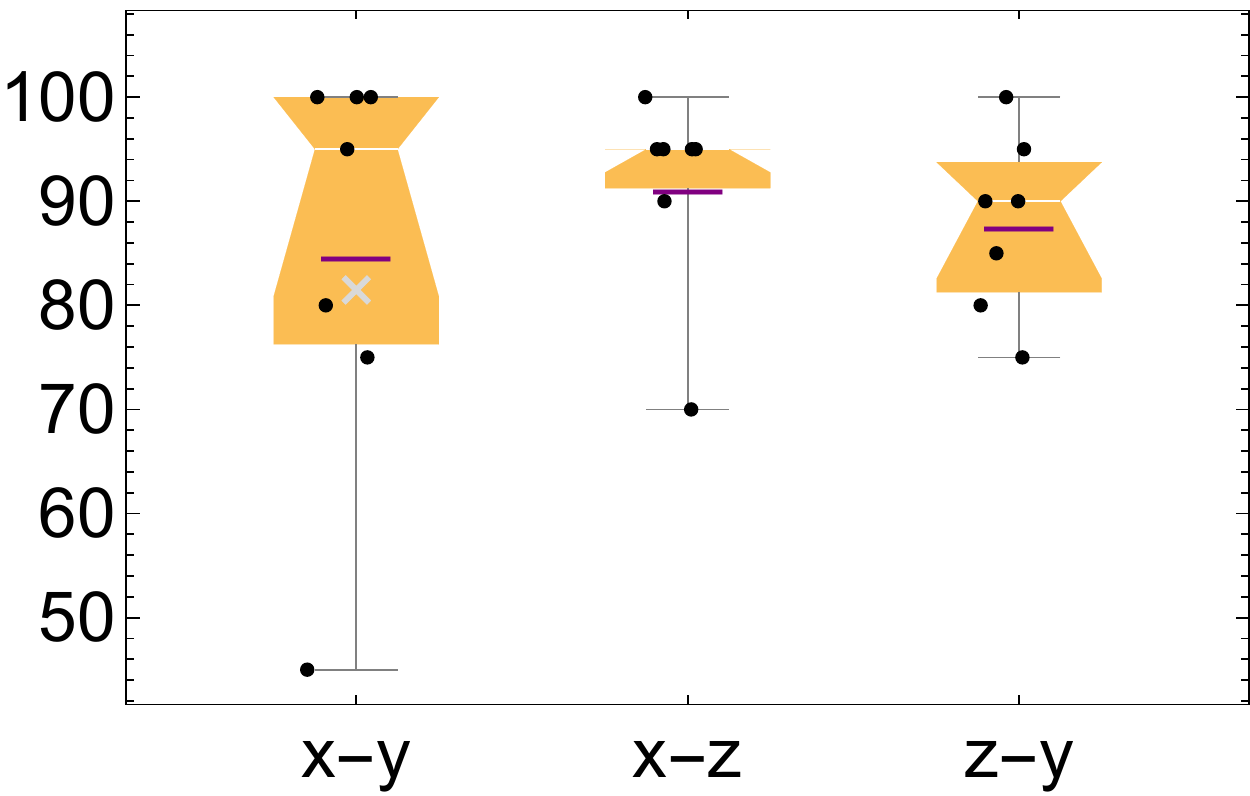}
\caption{Boxplot of either correctly identified field or its' direct neighbor per group, showing minimum and maximum (whiskers), median (white dash) and arithmetic mean (dark dash) and the performance of each individual participant (dots).}
\label{pic:boxn}
\end{figure}

\begin{figure}[thpb]
\centering
\includegraphics[width=0.45\textwidth]{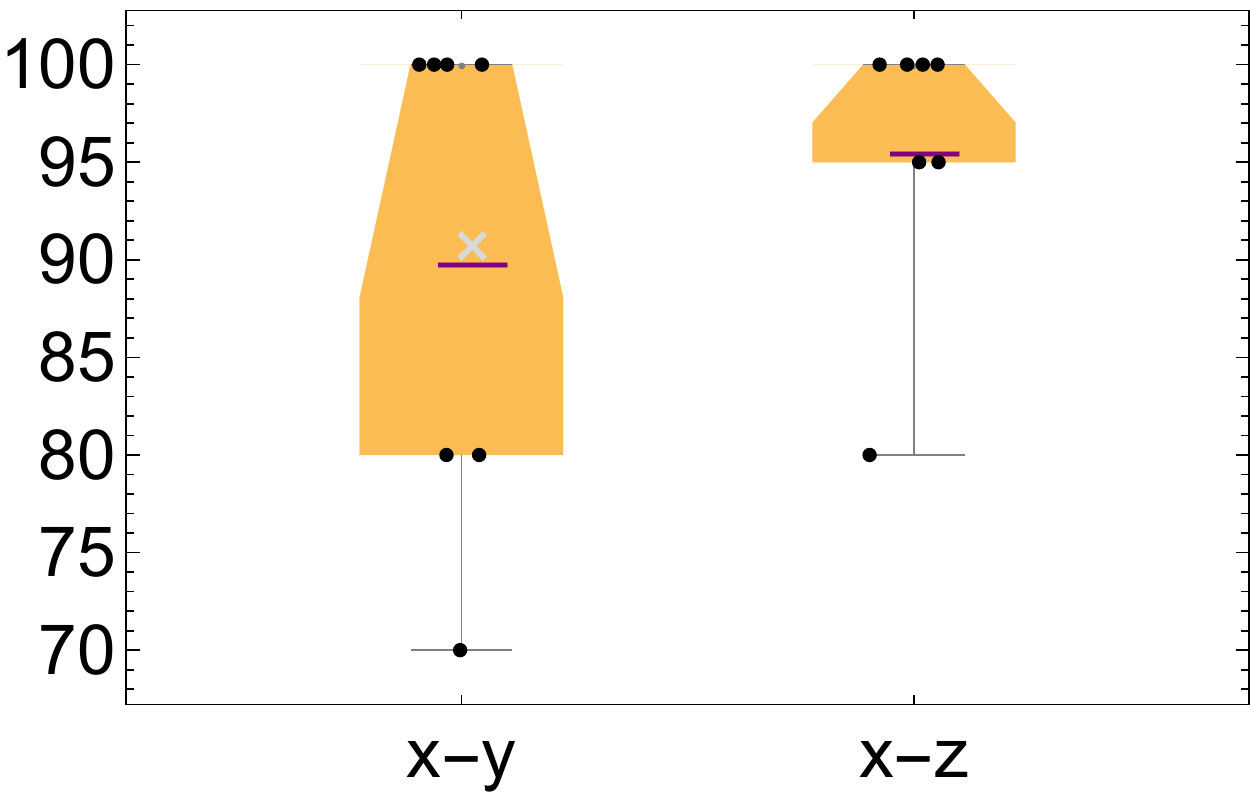}
\caption{Boxplot of correctly identified $x$-direction per group, showing minimum and maximum (whiskers), median (white dash) and arithmetic mean (dark dash),  the performance of each individual participant (dots) and the arithmetic mean (gray x) from our benchmark study \cite{poma,cars,dgm}.}
\label{pic:boxx}
\end{figure}

\begin{figure}[thpb]
\centering
\includegraphics[width=0.45\textwidth]{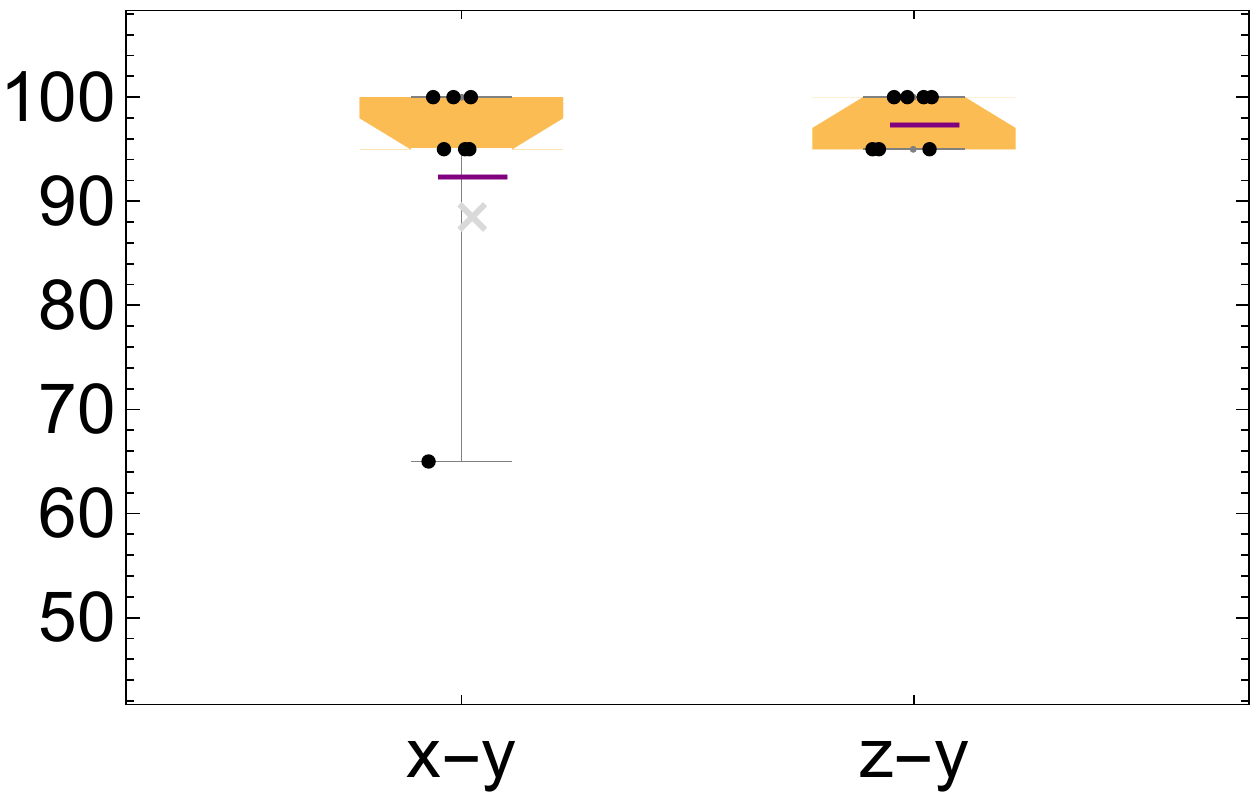}
\caption{Boxplot of correctly identified $y$-direction per group, showing minimum and maximum (whiskers), median (white dash) and arithmetic mean (dark dash), the performance of each individual participant (dots) and the arithmetic mean (gray x) from our benchmark study \cite{poma,cars,dgm}.}
\label{pic:boxy}
\end{figure}

\begin{figure}[thpb]
\centering
\includegraphics[width=0.45\textwidth]{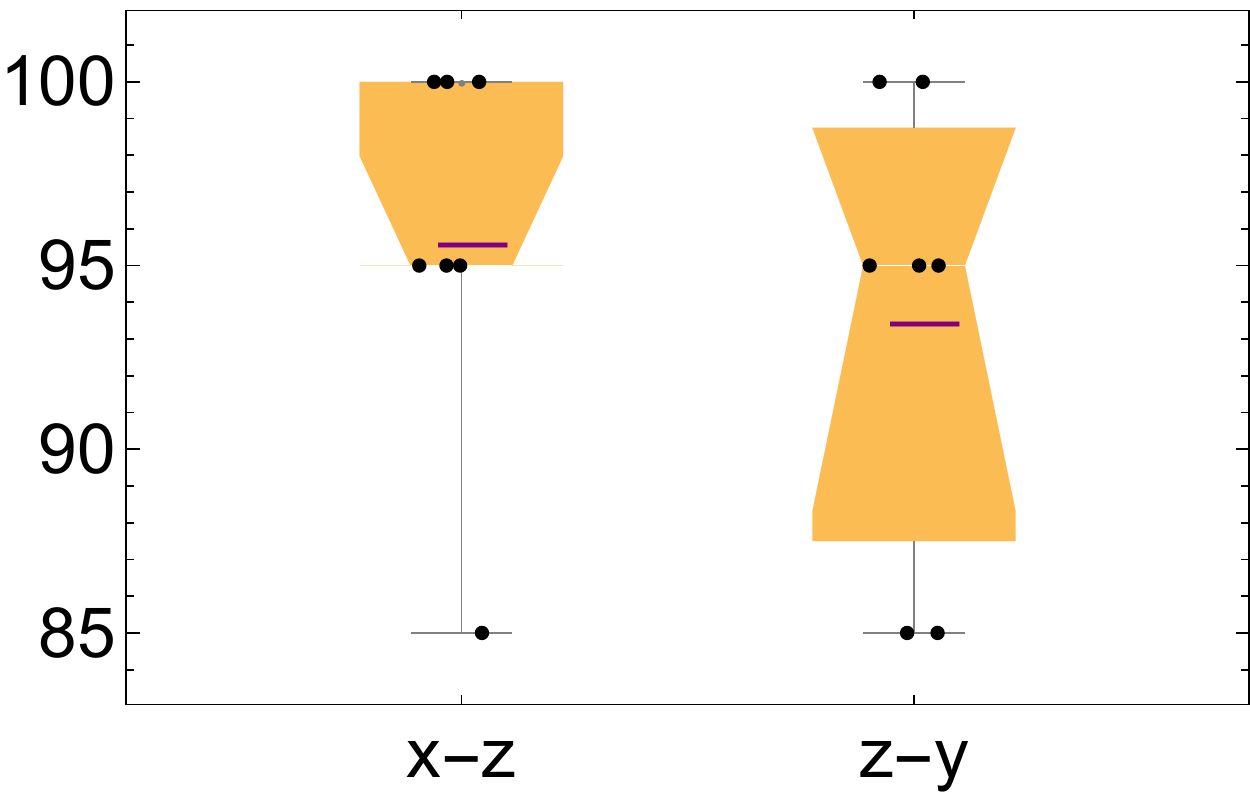}
\caption{Boxplot of correctly identified $z$-direction per group, showing minimum and maximum (whiskers), median (white dash) and arithmetic mean (dark dash) and the performance of each individual participant (dots).}
\label{pic:boxz}
\end{figure}

On average, it took participants roughly $7$ minutes to complete the experiment. The main results are shown in Figs. \ref{pic:boxhit} to \ref{pic:boxz}. The boxplots give details about user performances in the three scenarios. They show the score of each individual participant, the range, the $25$ and $75$ percentile, the median and the arithmetic mean value, and, where available, the results of our previous studies that serve as a benchmark \cite{poma,cars,dgm}. Overall, one can see that the results of all three groups are comparable in magnitude to the results of our previous study.

Fig. \ref{pic:boxhit} shows the hit rates of the three groups, which have a mean value between $51$\% and $64$\% and a median between $40$\% and $70$\%. Binomial tests indicated that the hit rate of every single participant was significantly higher than expected by chance (all $p$s$\leq0.001$). Accordingly, every participant was able to interpret the sonification. 
Fig. \ref{pic:boxq} shows the number of correct quadrants, having a mean value between $85$\% and $91$\% and a median between $90$\% and $95$\%. 
Fig. \ref{pic:boxn} shows how frequently the correct field or its direct neighbor was identified. Here, as for the quadrants, the mean value lay between $85$\% and $91$\% and the median between $90$\% and $95$\%. Figs.\ref{pic:boxx} to \ref{pic:boxz} show how often the $x$-, $y$-, or $z$-direction was identified correctly by the participants. Here, the arithmetic mean lay between $90$\% and $98$\%, the median between $95$\% and $100$\%. All these measures clearly show that the participants performed similarly well in all groups, and about as good as in our previous experiment, that had already been validated by an interactive experiment \cite{jmui}.

Of particular interest was to what extent previous experience with sonification had an influence on performance and also to what extent the different axis combinations were easier/harder to use than others. To investigate these two questions we proceeded as follows. First, we split participants into two groups based on experience: one group considered experienced (rating $\geq 4$) and the other considered inexperienced (rating $< 4$). We have both cases in each group, i.e., $4$ vs. $3$ in $x$-$y$, $5$ vs. $2$ in $x$-$z$, and $6$ vs. $1$ $z$-$y$. On average, the participants rated their previous experience with sonification with $2.14$ (median = $2$) .
Our evaluation measures  --- i.e., number of hits, correct quadrants, correctly identified field or neighbor, correct left/right direction and correct up/down direction --- exhibit relatively high correlation ($0.55\leq\rho\leq0.95$), as, for example a hit naturally comes along with a correct left/right and up/down direction, etc. We therefore carried out a Principal Component Analysis, to summarize the performance of the participants. Here, the first component explained $81.5$\% of the variance and the load of all measures on the component ranged between $0.79$ and $0.99$. Two-way analysis of variance (ANOVA) revealed no significant effect of previous experience and/or group on the results in terms of the first principal components($0.46<p<0.69$).

We have demonstrated already in \cite{poma,jmui}, that the $x$- and $y$-dimensions are orthogonal. Hence, we can conclude from the results of the statistical tests that also the new $z$-dimension is orthogonal to the previous ones, as the $x$-$z$- and $z$-$y$-group reveal no significant difference in terms of performance. This finding confirms that the new $z$ dimension is orthogonal to the $x$ and the $y$ dimension. The new $z$ dimension is equally well-combinable with the $x$ and the $y$ dimension. The fact that some participants have heard previous versions of the sonification before did not affect the results.

We observe that the $x$-$y$-group in this experiment performed better than in our previous study \cite{poma,dgm}. The main reason for this may be that we optimized the mapping based on the results of the previous study. Another reason may be the slight difference in the signal processing between \cite{acta} and \cite{icad2019}. Furthermore, letting the participants interactively explore the single and the combined audible dimensions may have improved their understanding of the perceptual auditory qualities and the psychoaocustic mapping principle. 




\begin{table*}[ptbh]
\begin{tabular}{|c||c|c|c|c||c|c|c|c||c|c|c|c||c|c|c|c|}
\hline
 ~& t1 & t2 & t3 & t4 & t5 & t6 & t7 & t8 & t9 & t19 & t11 & t12 & t13 & t14 & t15 & t16\\
\hline 
t1 & \cellcolor[gray]{0.2}$\color{white}80.$& \cellcolor[gray]{
0.9}$10.$& \cellcolor[gray]{0.9}$10.$& \cellcolor[gray]{
1.}$0.$& \cellcolor[gray]{1.}$0.$& \cellcolor[gray]{
1.}$0.$& \cellcolor[gray]{1.}$0.$& \cellcolor[gray]{
1.}$0.$& \cellcolor[gray]{1.}$0.$& \cellcolor[gray]{
1.}$0.$& \cellcolor[gray]{1.}$0.$& \cellcolor[gray]{
1.}$0.$& \cellcolor[gray]{1.}$0.$& \cellcolor[gray]{
1.}$0.$& \cellcolor[gray]{1.}$0.$& \cellcolor[gray]{1.}$0.$\\ t2 &
\cellcolor[gray]{0.667}$33.3$& \cellcolor[gray]{
0.556}$44.4$& \cellcolor[gray]{0.778}$22.2$& \cellcolor[gray]{
1.}$0.$& \cellcolor[gray]{1.}$0.$& \cellcolor[gray]{
1.}$0.$& \cellcolor[gray]{1.}$0.$& \cellcolor[gray]{
1.}$0.$& \cellcolor[gray]{1.}$0.$& \cellcolor[gray]{
1.}$0.$& \cellcolor[gray]{1.}$0.$& \cellcolor[gray]{
1.}$0.$& \cellcolor[gray]{1.}$0.$& \cellcolor[gray]{
1.}$0.$& \cellcolor[gray]{1.}$0.$& \cellcolor[gray]{1.}$0.$\\ t3 &
\cellcolor[gray]{0.8}$20.$& \cellcolor[gray]{0.9}$10.$& \cellcolor[gray]{0.4}$\color{white}60.$& \cellcolor[gray]{
1.}$0.$& \cellcolor[gray]{1.}$0.$& \cellcolor[gray]{
1.}$0.$& \cellcolor[gray]{0.9}$10.$& \cellcolor[gray]{
1.}$0.$& \cellcolor[gray]{1.}$0.$& \cellcolor[gray]{
1.}$0.$& \cellcolor[gray]{1.}$0.$& \cellcolor[gray]{
1.}$0.$& \cellcolor[gray]{1.}$0.$& \cellcolor[gray]{
1.}$0.$& \cellcolor[gray]{1.}$0.$& \cellcolor[gray]{1.}$0.$\\ t4 &
\cellcolor[gray]{1.}$0.$& \cellcolor[gray]{0.889}$11.1$& \cellcolor[gray]{0.556}$44.4$& \cellcolor[gray]{0.667}$33.3$& \cellcolor[gray]{1.}$0.$& \cellcolor[gray]{0.889}$11.1$& \cellcolor[gray]{
1.}$0.$& \cellcolor[gray]{1.}$0.$& \cellcolor[gray]{
1.}$0.$& \cellcolor[gray]{1.}$0.$& \cellcolor[gray]{
1.}$0.$& \cellcolor[gray]{1.}$0.$& \cellcolor[gray]{
1.}$0.$& \cellcolor[gray]{1.}$0.$& \cellcolor[gray]{
1.}$0.$& \cellcolor[gray]{1.}$0.$\\\hline t5 &
\cellcolor[gray]{1.}$0.$& \cellcolor[gray]{1.}$0.$& \cellcolor[gray]{
1.}$0.$& \cellcolor[gray]{1.}$0.$& \cellcolor[gray]{
0.333}$\color{white}66.7$& \cellcolor[gray]{
0.778}$22.2$& \cellcolor[gray]{1.}$0.$& \cellcolor[gray]{
1.}$0.$& \cellcolor[gray]{0.889}$11.1$& \cellcolor[gray]{
1.}$0.$& \cellcolor[gray]{1.}$0.$& \cellcolor[gray]{
1.}$0.$& \cellcolor[gray]{1.}$0.$& \cellcolor[gray]{
1.}$0.$& \cellcolor[gray]{1.}$0.$& \cellcolor[gray]{1.}$0.$\\ t6 &
\cellcolor[gray]{1.}$0.$& \cellcolor[gray]{0.9}$10.$& \cellcolor[gray]{1.}$0.$& \cellcolor[gray]{1.}$0.$& \cellcolor[gray]{
0.9}$10.$& \cellcolor[gray]{0.3}$\color{white}70.$& \cellcolor[gray]{0.9}$10.$& \cellcolor[gray]{1.}$0.$& \cellcolor[gray]{
1.}$0.$& \cellcolor[gray]{1.}$0.$& \cellcolor[gray]{
1.}$0.$& \cellcolor[gray]{1.}$0.$& \cellcolor[gray]{
1.}$0.$& \cellcolor[gray]{1.}$0.$& \cellcolor[gray]{
1.}$0.$& \cellcolor[gray]{1.}$0.$\\ t7 &
\cellcolor[gray]{1.}$0.$& \cellcolor[gray]{1.}$0.$& \cellcolor[gray]{
1.}$0.$& \cellcolor[gray]{1.}$0.$& \cellcolor[gray]{
0.571}$42.9$& \cellcolor[gray]{0.857}$14.3$& \cellcolor[gray]{
0.571}$42.9$& \cellcolor[gray]{1.}$0.$& \cellcolor[gray]{
1.}$0.$& \cellcolor[gray]{1.}$0.$& \cellcolor[gray]{
1.}$0.$& \cellcolor[gray]{1.}$0.$& \cellcolor[gray]{
1.}$0.$& \cellcolor[gray]{1.}$0.$& \cellcolor[gray]{
1.}$0.$& \cellcolor[gray]{1.}$0.$\\ t8 &
\cellcolor[gray]{0.9}$10.$& \cellcolor[gray]{1.}$0.$& \cellcolor[gray]{1.}$0.$& \cellcolor[gray]{1.}$0.$& \cellcolor[gray]{
0.8}$20.$& \cellcolor[gray]{1.}$0.$& \cellcolor[gray]{
0.6}$40.$& \cellcolor[gray]{0.7}$30.$& \cellcolor[gray]{
1.}$0.$& \cellcolor[gray]{1.}$0.$& \cellcolor[gray]{
1.}$0.$& \cellcolor[gray]{1.}$0.$& \cellcolor[gray]{
1.}$0.$& \cellcolor[gray]{1.}$0.$& \cellcolor[gray]{
1.}$0.$& \cellcolor[gray]{1.}$0.$\\\hline t9 &
\cellcolor[gray]{1.}$0.$& \cellcolor[gray]{0.889}$11.1$& \cellcolor[gray]{1.}$0.$& \cellcolor[gray]{1.}$0.$& \cellcolor[gray]{
0.889}$11.1$& \cellcolor[gray]{1.}$0.$& \cellcolor[gray]{
1.}$0.$& \cellcolor[gray]{1.}$0.$& \cellcolor[gray]{
0.556}$44.4$& \cellcolor[gray]{1.}$0.$& \cellcolor[gray]{
0.667}$33.3$& \cellcolor[gray]{1.}$0.$& \cellcolor[gray]{
1.}$0.$& \cellcolor[gray]{1.}$0.$& \cellcolor[gray]{
1.}$0.$& \cellcolor[gray]{1.}$0.$\\ t10 &
\cellcolor[gray]{0.875}$12.5$& \cellcolor[gray]{1.}$0.$& \cellcolor[gray]{1.}$0.$& \cellcolor[gray]{1.}$0.$& \cellcolor[gray]{
1.}$0.$& \cellcolor[gray]{0.875}$12.5$& \cellcolor[gray]{
1.}$0.$& \cellcolor[gray]{1.}$0.$& \cellcolor[gray]{
0.875}$12.5$& \cellcolor[gray]{0.625}$37.5$& \cellcolor[gray]{
1.}$0.$& \cellcolor[gray]{0.875}$12.5$& \cellcolor[gray]{
1.}$0.$& \cellcolor[gray]{1.}$0.$& \cellcolor[gray]{
1.}$0.$& \cellcolor[gray]{0.875}$12.5$\\ t11 &
\cellcolor[gray]{1.}$0.$& \cellcolor[gray]{0.875}$12.5$& \cellcolor[gray]{1.}$0.$& \cellcolor[gray]{1.}$0.$& \cellcolor[gray]{
1.}$0.$& \cellcolor[gray]{1.}$0.$& \cellcolor[gray]{
0.875}$12.5$& \cellcolor[gray]{1.}$0.$& \cellcolor[gray]{
0.875}$12.5$& \cellcolor[gray]{1.}$0.$& \cellcolor[gray]{
0.375}$\color{white}62.5$& \cellcolor[gray]{1.}$0.$& \cellcolor[gray]{1.}$0.$& \cellcolor[gray]{1.}$0.$& \cellcolor[gray]{
1.}$0.$& \cellcolor[gray]{1.}$0.$\\ t12 &
\cellcolor[gray]{1.}$0.$& \cellcolor[gray]{1.}$0.$& \cellcolor[gray]{
1.}$0.$& \cellcolor[gray]{1.}$0.$& \cellcolor[gray]{
1.}$0.$& \cellcolor[gray]{1.}$0.$& \cellcolor[gray]{
1.}$0.$& \cellcolor[gray]{1.}$0.$& \cellcolor[gray]{
1.}$0.$& \cellcolor[gray]{1.}$0.$& \cellcolor[gray]{
0.625}$37.5$& \cellcolor[gray]{0.625}$37.5$& \cellcolor[gray]{
1.}$0.$& \cellcolor[gray]{0.875}$12.5$& \cellcolor[gray]{
0.875}$12.5$& \cellcolor[gray]{1.}$0.$\\\hline t13 &
\cellcolor[gray]{1.}$0.$& \cellcolor[gray]{0.857}$14.3$& \cellcolor[gray]{1.}$0.$& \cellcolor[gray]{1.}$0.$& \cellcolor[gray]{
1.}$0.$& \cellcolor[gray]{1.}$0.$& \cellcolor[gray]{
1.}$0.$& \cellcolor[gray]{1.}$0.$& \cellcolor[gray]{
1.}$0.$& \cellcolor[gray]{1.}$0.$& \cellcolor[gray]{
1.}$0.$& \cellcolor[gray]{1.}$0.$& \cellcolor[gray]{
0.286}$\color{white}71.4$& \cellcolor[gray]{
0.857}$14.3$& \cellcolor[gray]{1.}$0.$& \cellcolor[gray]{
1.}$0.$\\ t14 &
\cellcolor[gray]{1.}$0.$& \cellcolor[gray]{1.}$0.$& \cellcolor[gray]{
1.}$0.$& \cellcolor[gray]{1.}$0.$& \cellcolor[gray]{
1.}$0.$& \cellcolor[gray]{1.}$0.$& \cellcolor[gray]{
1.}$0.$& \cellcolor[gray]{1.}$0.$& \cellcolor[gray]{
1.}$0.$& \cellcolor[gray]{1.}$0.$& \cellcolor[gray]{
0.714}$28.6$& \cellcolor[gray]{1.}$0.$& \cellcolor[gray]{
1.}$0.$& \cellcolor[gray]{0.429}$\color{white}57.1$& \cellcolor[gray]{0.857}$14.3$& \cellcolor[gray]{1.}$0.$\\ t15 &
\cellcolor[gray]{0.9}$10.$& \cellcolor[gray]{1.}$0.$& \cellcolor[gray]{1.}$0.$& \cellcolor[gray]{0.9}$10.$& \cellcolor[gray]{
1.}$0.$& \cellcolor[gray]{1.}$0.$& \cellcolor[gray]{
1.}$0.$& \cellcolor[gray]{1.}$0.$& \cellcolor[gray]{
0.9}$10.$& \cellcolor[gray]{1.}$0.$& \cellcolor[gray]{
1.}$0.$& \cellcolor[gray]{1.}$0.$& \cellcolor[gray]{
0.9}$10.$& \cellcolor[gray]{0.9}$10.$& \cellcolor[gray]{
0.5}$50.$& \cellcolor[gray]{1.}$0.$\\ t16 &
\cellcolor[gray]{1.}$0.$& \cellcolor[gray]{1.}$0.$& \cellcolor[gray]{
1.}$0.$& \cellcolor[gray]{1.}$0.$& \cellcolor[gray]{
1.}$0.$& \cellcolor[gray]{1.}$0.$& \cellcolor[gray]{
1.}$0.$& \cellcolor[gray]{1.}$0.$& \cellcolor[gray]{
1.}$0.$& \cellcolor[gray]{0.889}$11.1$& \cellcolor[gray]{
1.}$0.$& \cellcolor[gray]{1.}$0.$& \cellcolor[gray]{
0.889}$11.1$& \cellcolor[gray]{0.778}$22.2$& \cellcolor[gray]{
0.889}$11.1$& \cellcolor[gray]{0.556}$44.4$\\\hline
\end{tabular}
\caption{Confusion matrix (in percentages) for the $x$-$y$-plane. The rows represent the sonified targets, the columns the selected field. The values indicate how often each field has been clicked when either of the targets was sonified.
}
\label{tab:confusion1}
\end{table*}

\begin{table*}[ptbh]
\begin{tabular}{|c||c|c|c|c||c|c|c|c||c|c|c|c||c|c|c|c|}
\hline
 & t1 & t2 & t3 & t4 & t5 & t6 & t7 & t8 & t9 & t19 & t11 & t12 & t13 & t14 & t15 & t16\\
\hline 
t1 & \cellcolor[gray]{0.091}$\color{white}90.9$& \cellcolor[gray]{
1.}$0.$& \cellcolor[gray]{0.909}$9.1$& \cellcolor[gray]{
1.}$0.$& \cellcolor[gray]{1.}$0.$& \cellcolor[gray]{
1.}$0.$& \cellcolor[gray]{1.}$0.$& \cellcolor[gray]{
1.}$0.$& \cellcolor[gray]{1.}$0.$& \cellcolor[gray]{
1.}$0.$& \cellcolor[gray]{1.}$0.$& \cellcolor[gray]{
1.}$0.$& \cellcolor[gray]{1.}$0.$& \cellcolor[gray]{
1.}$0.$& \cellcolor[gray]{1.}$0.$& \cellcolor[gray]{1.}$0.$\\ t2 & 
\cellcolor[gray]{0.875}$12.5$& \cellcolor[gray]{0.25}$\color{white}
75.$& \cellcolor[gray]{1.}$0.$& \cellcolor[gray]{
1.}$0.$& \cellcolor[gray]{1.}$0.$& \cellcolor[gray]{
1.}$0.$& \cellcolor[gray]{0.875}$12.5$& \cellcolor[gray]{
1.}$0.$& \cellcolor[gray]{1.}$0.$& \cellcolor[gray]{
1.}$0.$& \cellcolor[gray]{1.}$0.$& \cellcolor[gray]{
1.}$0.$& \cellcolor[gray]{1.}$0.$& \cellcolor[gray]{
1.}$0.$& \cellcolor[gray]{1.}$0.$& \cellcolor[gray]{1.}$0.$\\ t3 & 
\cellcolor[gray]{0.778}$22.2$& \cellcolor[gray]{
0.889}$11.1$& \cellcolor[gray]{0.333}$\color{white}
66.7$& \cellcolor[gray]{1.}$0.$& \cellcolor[gray]{
1.}$0.$& \cellcolor[gray]{1.}$0.$& \cellcolor[gray]{
1.}$0.$& \cellcolor[gray]{1.}$0.$& \cellcolor[gray]{
1.}$0.$& \cellcolor[gray]{1.}$0.$& \cellcolor[gray]{
1.}$0.$& \cellcolor[gray]{1.}$0.$& \cellcolor[gray]{
1.}$0.$& \cellcolor[gray]{1.}$0.$& \cellcolor[gray]{
1.}$0.$& \cellcolor[gray]{1.}$0.$\\ t4 & 
\cellcolor[gray]{1.}$0.$& \cellcolor[gray]{0.6}$40.$& \cellcolor[gray]{0.8}$20.$& \cellcolor[gray]{0.7}$30.$& \cellcolor[gray]{
1.}$0.$& \cellcolor[gray]{1.}$0.$& \cellcolor[gray]{
1.}$0.$& \cellcolor[gray]{0.9}$10.$& \cellcolor[gray]{
1.}$0.$& \cellcolor[gray]{1.}$0.$& \cellcolor[gray]{
1.}$0.$& \cellcolor[gray]{1.}$0.$& \cellcolor[gray]{
1.}$0.$& \cellcolor[gray]{1.}$0.$& \cellcolor[gray]{
1.}$0.$& \cellcolor[gray]{1.}$0.$\\\hline t5 & 
\cellcolor[gray]{1.}$0.$& \cellcolor[gray]{1.}$0.$& \cellcolor[gray]{
1.}$0.$& \cellcolor[gray]{1.}$0.$& \cellcolor[gray]{
0.125}$\color{white}87.5$& \cellcolor[gray]{1.}$0.$& \cellcolor[gray]{1.}$0.$& \cellcolor[gray]{0.875}$12.5$& \cellcolor[gray]{
1.}$0.$& \cellcolor[gray]{1.}$0.$& \cellcolor[gray]{
1.}$0.$& \cellcolor[gray]{1.}$0.$& \cellcolor[gray]{
1.}$0.$& \cellcolor[gray]{1.}$0.$& \cellcolor[gray]{
1.}$0.$& \cellcolor[gray]{1.}$0.$\\ t6 & 
\cellcolor[gray]{1.}$0.$& \cellcolor[gray]{1.}$0.$& \cellcolor[gray]{
1.}$0.$& \cellcolor[gray]{1.}$0.$& \cellcolor[gray]{0.4}$\color{
white}60.$& \cellcolor[gray]{0.7}$30.$& \cellcolor[gray]{
1.}$0.$& \cellcolor[gray]{1.}$0.$& \cellcolor[gray]{
1.}$0.$& \cellcolor[gray]{1.}$0.$& \cellcolor[gray]{
0.9}$10.$& \cellcolor[gray]{1.}$0.$& \cellcolor[gray]{
1.}$0.$& \cellcolor[gray]{1.}$0.$& \cellcolor[gray]{
1.}$0.$& \cellcolor[gray]{1.}$0.$\\ t7 & 
\cellcolor[gray]{1.}$0.$& \cellcolor[gray]{1.}$0.$& \cellcolor[gray]{
1.}$0.$& \cellcolor[gray]{1.}$0.$& \cellcolor[gray]{
1.}$0.$& \cellcolor[gray]{1.}$0.$& \cellcolor[gray]{
0.25}$\color{white}75.$& \cellcolor[gray]{
0.875}$12.5$& \cellcolor[gray]{1.}$0.$& \cellcolor[gray]{
0.875}$12.5$& \cellcolor[gray]{1.}$0.$& \cellcolor[gray]{
1.}$0.$& \cellcolor[gray]{1.}$0.$& \cellcolor[gray]{
1.}$0.$& \cellcolor[gray]{1.}$0.$& \cellcolor[gray]{1.}$0.$\\ t8 & 
\cellcolor[gray]{1.}$0.$& \cellcolor[gray]{1.}$0.$& \cellcolor[gray]{
1.}$0.$& \cellcolor[gray]{1.}$0.$& \cellcolor[gray]{
0.889}$11.1$& \cellcolor[gray]{0.889}$11.1$& \cellcolor[gray]{
0.889}$11.1$& \cellcolor[gray]{0.333}$\color{white}
66.7$& \cellcolor[gray]{1.}$0.$& \cellcolor[gray]{
1.}$0.$& \cellcolor[gray]{1.}$0.$& \cellcolor[gray]{
1.}$0.$& \cellcolor[gray]{1.}$0.$& \cellcolor[gray]{
1.}$0.$& \cellcolor[gray]{1.}$0.$& \cellcolor[gray]{1.}$0.$\\\hline t9 & 
\cellcolor[gray]{1.}$0.$& \cellcolor[gray]{1.}$0.$& \cellcolor[gray]{
1.}$0.$& \cellcolor[gray]{1.}$0.$& \cellcolor[gray]{
0.909}$9.1$& \cellcolor[gray]{1.}$0.$& \cellcolor[gray]{
1.}$0.$& \cellcolor[gray]{1.}$0.$& \cellcolor[gray]{
0.364}$\color{white}63.6$& \cellcolor[gray]{1.}$0.$& \cellcolor[gray]{0.727}$27.3$& \cellcolor[gray]{1.}$0.$& \cellcolor[gray]{
1.}$0.$& \cellcolor[gray]{1.}$0.$& \cellcolor[gray]{
1.}$0.$& \cellcolor[gray]{1.}$0.$\\ t10 & 
\cellcolor[gray]{1.}$0.$& \cellcolor[gray]{1.}$0.$& \cellcolor[gray]{
1.}$0.$& \cellcolor[gray]{1.}$0.$& \cellcolor[gray]{
1.}$0.$& \cellcolor[gray]{1.}$0.$& \cellcolor[gray]{
1.}$0.$& \cellcolor[gray]{1.}$0.$& \cellcolor[gray]{
1.}$0.$& \cellcolor[gray]{0.286}$\color{white}71.4$& \cellcolor[gray]{1.}$0.$& \cellcolor[gray]{0.857}$14.3$& \cellcolor[gray]{
1.}$0.$& \cellcolor[gray]{1.}$0.$& \cellcolor[gray]{
0.857}$14.3$& \cellcolor[gray]{1.}$0.$\\ t11 & 
\cellcolor[gray]{1.}$0.$& \cellcolor[gray]{1.}$0.$& \cellcolor[gray]{
1.}$0.$& \cellcolor[gray]{1.}$0.$& \cellcolor[gray]{
1.}$0.$& \cellcolor[gray]{1.}$0.$& \cellcolor[gray]{
1.}$0.$& \cellcolor[gray]{1.}$0.$& \cellcolor[gray]{
0.75}$25.$& \cellcolor[gray]{1.}$0.$& \cellcolor[gray]{
0.25}$\color{white}75.$& \cellcolor[gray]{1.}$0.$& \cellcolor[gray]{1.}$0.$& \cellcolor[gray]{1.}$0.$& \cellcolor[gray]{
1.}$0.$& \cellcolor[gray]{1.}$0.$\\ t12 & 
\cellcolor[gray]{1.}$0.$& \cellcolor[gray]{1.}$0.$& \cellcolor[gray]{
1.}$0.$& \cellcolor[gray]{1.}$0.$& \cellcolor[gray]{
1.}$0.$& \cellcolor[gray]{1.}$0.$& \cellcolor[gray]{
1.}$0.$& \cellcolor[gray]{1.}$0.$& \cellcolor[gray]{
1.}$0.$& \cellcolor[gray]{0.875}$12.5$& \cellcolor[gray]{
0.875}$12.5$& \cellcolor[gray]{0.375}$\color{white}
62.5$& \cellcolor[gray]{1.}$0.$& \cellcolor[gray]{
0.875}$12.5$& \cellcolor[gray]{1.}$0.$& \cellcolor[gray]{
1.}$0.$\\\hline t13 & 
\cellcolor[gray]{0.875}$12.5$& \cellcolor[gray]{1.}$0.$& \cellcolor[gray]{1.}$0.$& \cellcolor[gray]{1.}$0.$& \cellcolor[gray]{
1.}$0.$& \cellcolor[gray]{1.}$0.$& \cellcolor[gray]{
1.}$0.$& \cellcolor[gray]{1.}$0.$& \cellcolor[gray]{
1.}$0.$& \cellcolor[gray]{1.}$0.$& \cellcolor[gray]{
1.}$0.$& \cellcolor[gray]{1.}$0.$& \cellcolor[gray]{
0.625}$37.5$& \cellcolor[gray]{0.5}$50.$& \cellcolor[gray]{
1.}$0.$& \cellcolor[gray]{1.}$0.$\\ t14 & 
\cellcolor[gray]{1.}$0.$& \cellcolor[gray]{1.}$0.$& \cellcolor[gray]{
1.}$0.$& \cellcolor[gray]{1.}$0.$& \cellcolor[gray]{
1.}$0.$& \cellcolor[gray]{1.}$0.$& \cellcolor[gray]{
1.}$0.$& \cellcolor[gray]{1.}$0.$& \cellcolor[gray]{
1.}$0.$& \cellcolor[gray]{1.}$0.$& \cellcolor[gray]{
0.889}$11.1$& \cellcolor[gray]{1.}$0.$& \cellcolor[gray]{
1.}$0.$& \cellcolor[gray]{0.444}$\color{white}55.6$& \cellcolor[gray]{0.889}$11.1$& \cellcolor[gray]{0.778}$22.2$\\ t15 & 
\cellcolor[gray]{1.}$0.$& \cellcolor[gray]{0.75}$25.$& \cellcolor[gray]{1.}$0.$& \cellcolor[gray]{1.}$0.$& \cellcolor[gray]{
1.}$0.$& \cellcolor[gray]{1.}$0.$& \cellcolor[gray]{
1.}$0.$& \cellcolor[gray]{1.}$0.$& \cellcolor[gray]{
1.}$0.$& \cellcolor[gray]{1.}$0.$& \cellcolor[gray]{
1.}$0.$& \cellcolor[gray]{1.}$0.$& \cellcolor[gray]{
1.}$0.$& \cellcolor[gray]{1.}$0.$& \cellcolor[gray]{
0.375}$\color{white}62.5$& \cellcolor[gray]{0.875}$12.5$\\ t16 & 
\cellcolor[gray]{1.}$0.$& \cellcolor[gray]{1.}$0.$& \cellcolor[gray]{
1.}$0.$& \cellcolor[gray]{1.}$0.$& \cellcolor[gray]{
1.}$0.$& \cellcolor[gray]{1.}$0.$& \cellcolor[gray]{
1.}$0.$& \cellcolor[gray]{1.}$0.$& \cellcolor[gray]{
1.}$0.$& \cellcolor[gray]{1.}$0.$& \cellcolor[gray]{
1.}$0.$& \cellcolor[gray]{0.875}$12.5$& \cellcolor[gray]{
1.}$0.$& \cellcolor[gray]{1.}$0.$& \cellcolor[gray]{
0.875}$12.5$& \cellcolor[gray]{0.25}$\color{white}75.$\\\hline
\end{tabular}
\caption{Confusion matrix (in percentages) for the $x$-$z$-plane. The rows represent the sonified targets, the columns the selected field. The values indicate how often each field has been clicked when either of the targets was sonified.
}
\label{tab:confusion2}
\end{table*}

\begin{table*}[ptbh]
\begin{tabular}{|c||c|c|c|c||c|c|c|c||c|c|c|c||c|c|c|c|}
\hline
 & t1 & t2 & t3 & t4 & t5 & t6 & t7 & t8 & t9 & t19 & t11 & t12 & t13 & t14 & t15 & t16\\
\hline 
t1 & \cellcolor[gray]{0.444}$\color{white}55.6$& \cellcolor[gray]{
0.889}$11.1$& \cellcolor[gray]{1.}$0.$& \cellcolor[gray]{
1.}$0.$& \cellcolor[gray]{0.889}$11.1$& \cellcolor[gray]{
0.778}$22.2$& \cellcolor[gray]{1.}$0.$& \cellcolor[gray]{
1.}$0.$& \cellcolor[gray]{1.}$0.$& \cellcolor[gray]{
1.}$0.$& \cellcolor[gray]{1.}$0.$& \cellcolor[gray]{1.}$0.$& \cellcolor[gray]{1.}$0.$& \cellcolor[gray]{
1.}$0.$& \cellcolor[gray]{1.}$0.$& \cellcolor[gray]{1.}$0.$ \\ t2 &
\cellcolor[gray]{0.545}$45.5$& \cellcolor[gray]{
0.636}$36.4$& \cellcolor[gray]{1.}$0.$& \cellcolor[gray]{
0.909}$9.1$& \cellcolor[gray]{0.909}$9.1$& \cellcolor[gray]{
1.}$0.$& \cellcolor[gray]{1.}$0.$& \cellcolor[gray]{
1.}$0.$& \cellcolor[gray]{1.}$0.$& \cellcolor[gray]{
1.}$0.$& \cellcolor[gray]{1.}$0.$& \cellcolor[gray]{
1.}$0.$& \cellcolor[gray]{1.}$0.$& \cellcolor[gray]{
1.}$0.$& \cellcolor[gray]{1.}$0.$& \cellcolor[gray]{1.}$0.$ \\ t3 &
\cellcolor[gray]{0.875}$12.5$& \cellcolor[gray]{1.}$0.$& \cellcolor[gray]{0.25}$\color{white}75.$& \cellcolor[gray]{
1.}$0.$& \cellcolor[gray]{1.}$0.$& \cellcolor[gray]{
0.875}$12.5$& \cellcolor[gray]{1.}$0.$& \cellcolor[gray]{
1.}$0.$& \cellcolor[gray]{1.}$0.$& \cellcolor[gray]{
1.}$0.$& \cellcolor[gray]{1.}$0.$& \cellcolor[gray]{
1.}$0.$& \cellcolor[gray]{1.}$0.$& \cellcolor[gray]{
1.}$0.$& \cellcolor[gray]{1.}$0.$& \cellcolor[gray]{1.}$0.$ \\  t4 &
\cellcolor[gray]{1.}$0.$& \cellcolor[gray]{0.714}$28.6$& \cellcolor[gray]{0.714}$28.6$& \cellcolor[gray]{0.714}$28.6$& \cellcolor[gray]{1.}$0.$& \cellcolor[gray]{0.857}$14.3$& \cellcolor[gray]{
1.}$0.$& \cellcolor[gray]{1.}$0.$& \cellcolor[gray]{
1.}$0.$& \cellcolor[gray]{1.}$0.$& \cellcolor[gray]{
1.}$0.$& \cellcolor[gray]{1.}$0.$& \cellcolor[gray]{
1.}$0.$& \cellcolor[gray]{1.}$0.$& \cellcolor[gray]{
1.}$0.$& \cellcolor[gray]{1.}$0.$ \\\hline t5 &
\cellcolor[gray]{1.}$0.$& \cellcolor[gray]{1.}$0.$& \cellcolor[gray]{
1.}$0.$& \cellcolor[gray]{1.}$0.$& \cellcolor[gray]{
0.375}$\color{white}62.5$& \cellcolor[gray]{
0.75}$25.$& \cellcolor[gray]{0.875}$12.5$& \cellcolor[gray]{
1.}$0.$& \cellcolor[gray]{1.}$0.$& \cellcolor[gray]{
1.}$0.$& \cellcolor[gray]{1.}$0.$& \cellcolor[gray]{
1.}$0.$& \cellcolor[gray]{1.}$0.$& \cellcolor[gray]{
1.}$0.$& \cellcolor[gray]{1.}$0.$& \cellcolor[gray]{1.}$0.$ \\ t6 &
\cellcolor[gray]{1.}$0.$& \cellcolor[gray]{1.}$0.$& \cellcolor[gray]{
1.}$0.$& \cellcolor[gray]{1.}$0.$& \cellcolor[gray]{
1.}$0.$& \cellcolor[gray]{0.222}$\color{white}77.8$& \cellcolor[gray]{1.}$0.$& \cellcolor[gray]{0.778}$22.2$& \cellcolor[gray]{
1.}$0.$& \cellcolor[gray]{1.}$0.$& \cellcolor[gray]{
1.}$0.$& \cellcolor[gray]{1.}$0.$& \cellcolor[gray]{
1.}$0.$& \cellcolor[gray]{1.}$0.$& \cellcolor[gray]{
1.}$0.$& \cellcolor[gray]{1.}$0.$ \\ t7 &
\cellcolor[gray]{1.}$0.$& \cellcolor[gray]{1.}$0.$& \cellcolor[gray]{
1.}$0.$& \cellcolor[gray]{1.}$0.$& \cellcolor[gray]{
1.}$0.$& \cellcolor[gray]{1.}$0.$& \cellcolor[gray]{
0.222}$\color{white}77.8$& \cellcolor[gray]{
0.778}$22.2$& \cellcolor[gray]{1.}$0.$& \cellcolor[gray]{
1.}$0.$& \cellcolor[gray]{1.}$0.$& \cellcolor[gray]{
1.}$0.$& \cellcolor[gray]{1.}$0.$& \cellcolor[gray]{
1.}$0.$& \cellcolor[gray]{1.}$0.$& \cellcolor[gray]{1.}$0.$ \\ t8 &
\cellcolor[gray]{1.}$0.$& \cellcolor[gray]{1.}$0.$& \cellcolor[gray]{
1.}$0.$& \cellcolor[gray]{1.}$0.$& \cellcolor[gray]{
1.}$0.$& \cellcolor[gray]{0.875}$12.5$& \cellcolor[gray]{
0.875}$12.5$& \cellcolor[gray]{0.375}$\color{white}
62.5$& \cellcolor[gray]{1.}$0.$& \cellcolor[gray]{
0.875}$12.5$& \cellcolor[gray]{1.}$0.$& \cellcolor[gray]{
1.}$0.$& \cellcolor[gray]{1.}$0.$& \cellcolor[gray]{
1.}$0.$& \cellcolor[gray]{1.}$0.$& \cellcolor[gray]{1.}$0.$ \\\hline  t9 &
\cellcolor[gray]{1.}$0.$& \cellcolor[gray]{1.}$0.$& \cellcolor[gray]{
1.}$0.$& \cellcolor[gray]{1.}$0.$& \cellcolor[gray]{
1.}$0.$& \cellcolor[gray]{1.}$0.$& \cellcolor[gray]{
1.}$0.$& \cellcolor[gray]{1.}$0.$& \cellcolor[gray]{
0.7}$30.$& \cellcolor[gray]{0.8}$20.$& \cellcolor[gray]{
0.7}$30.$& \cellcolor[gray]{0.9}$10.$& \cellcolor[gray]{
1.}$0.$& \cellcolor[gray]{0.9}$10.$& \cellcolor[gray]{
1.}$0.$& \cellcolor[gray]{1.}$0.$ \\ t10 &
\cellcolor[gray]{1.}$0.$& \cellcolor[gray]{1.}$0.$& \cellcolor[gray]{
1.}$0.$& \cellcolor[gray]{1.}$0.$& \cellcolor[gray]{
1.}$0.$& \cellcolor[gray]{1.}$0.$& \cellcolor[gray]{
1.}$0.$& \cellcolor[gray]{1.}$0.$& \cellcolor[gray]{
1.}$0.$& \cellcolor[gray]{0.7}$30.$& \cellcolor[gray]{
1.}$0.$& \cellcolor[gray]{0.3}$\color{white}70.$& \cellcolor[gray]{1.}$0.$& \cellcolor[gray]{1.}$0.$& \cellcolor[gray]{
1.}$0.$& \cellcolor[gray]{1.}$0.$ \\ t11 &
\cellcolor[gray]{1.}$0.$& \cellcolor[gray]{1.}$0.$& \cellcolor[gray]{
1.}$0.$& \cellcolor[gray]{1.}$0.$& \cellcolor[gray]{
1.}$0.$& \cellcolor[gray]{1.}$0.$& \cellcolor[gray]{
1.}$0.$& \cellcolor[gray]{1.}$0.$& \cellcolor[gray]{
1.}$0.$& \cellcolor[gray]{1.}$0.$& \cellcolor[gray]{
0.545}$45.5$& \cellcolor[gray]{0.727}$27.3$& \cellcolor[gray]{
1.}$0.$& \cellcolor[gray]{1.}$0.$& \cellcolor[gray]{
0.909}$9.1$& \cellcolor[gray]{0.818}$18.2$ \\ t12 &
\cellcolor[gray]{1.}$0.$& \cellcolor[gray]{1.}$0.$& \cellcolor[gray]{
1.}$0.$& \cellcolor[gray]{1.}$0.$& \cellcolor[gray]{
1.}$0.$& \cellcolor[gray]{0.857}$14.3$& \cellcolor[gray]{
1.}$0.$& \cellcolor[gray]{0.857}$14.3$& \cellcolor[gray]{
1.}$0.$& \cellcolor[gray]{1.}$0.$& \cellcolor[gray]{
1.}$0.$& \cellcolor[gray]{0.286}$\color{white}71.4$& \cellcolor[gray]{1.}$0.$& \cellcolor[gray]{1.}$0.$& \cellcolor[gray]{
1.}$0.$& \cellcolor[gray]{1.}$0.$ \\\hline t13 &
\cellcolor[gray]{1.}$0.$& \cellcolor[gray]{1.}$0.$& \cellcolor[gray]{
1.}$0.$& \cellcolor[gray]{1.}$0.$& \cellcolor[gray]{
1.}$0.$& \cellcolor[gray]{1.}$0.$& \cellcolor[gray]{
1.}$0.$& \cellcolor[gray]{1.}$0.$& \cellcolor[gray]{
1.}$0.$& \cellcolor[gray]{1.}$0.$& \cellcolor[gray]{
1.}$0.$& \cellcolor[gray]{1.}$0.$& \cellcolor[gray]{
0.571}$42.9$& \cellcolor[gray]{0.571}$42.9$& \cellcolor[gray]{
1.}$0.$& \cellcolor[gray]{0.857}$14.3$ \\ t14 &
\cellcolor[gray]{1.}$0.$& \cellcolor[gray]{1.}$0.$& \cellcolor[gray]{
1.}$0.$& \cellcolor[gray]{1.}$0.$& \cellcolor[gray]{
1.}$0.$& \cellcolor[gray]{1.}$0.$& \cellcolor[gray]{
1.}$0.$& \cellcolor[gray]{1.}$0.$& \cellcolor[gray]{
1.}$0.$& \cellcolor[gray]{1.}$0.$& \cellcolor[gray]{
0.875}$12.5$& \cellcolor[gray]{1.}$0.$& \cellcolor[gray]{
0.875}$12.5$& \cellcolor[gray]{0.75}$25.$& \cellcolor[gray]{
0.75}$25.$& \cellcolor[gray]{0.75}$25.$ \\ t15 &
\cellcolor[gray]{1.}$0.$& \cellcolor[gray]{1.}$0.$& \cellcolor[gray]{
1.}$0.$& \cellcolor[gray]{1.}$0.$& \cellcolor[gray]{
1.}$0.$& \cellcolor[gray]{1.}$0.$& \cellcolor[gray]{
1.}$0.$& \cellcolor[gray]{1.}$0.$& \cellcolor[gray]{
1.}$0.$& \cellcolor[gray]{1.}$0.$& \cellcolor[gray]{
1.}$0.$& \cellcolor[gray]{1.}$0.$& \cellcolor[gray]{
0.9}$10.$& \cellcolor[gray]{0.9}$10.$& \cellcolor[gray]{
0.4}$\color{white}60.$& \cellcolor[gray]{0.8}$20.$ \\ t16 &
\cellcolor[gray]{1.}$0.$& \cellcolor[gray]{1.}$0.$& \cellcolor[gray]{
1.}$0.$& \cellcolor[gray]{1.}$0.$& \cellcolor[gray]{
1.}$0.$& \cellcolor[gray]{1.}$0.$& \cellcolor[gray]{
1.}$0.$& \cellcolor[gray]{1.}$0.$& \cellcolor[gray]{
1.}$0.$& \cellcolor[gray]{1.}$0.$& \cellcolor[gray]{
1.}$0.$& \cellcolor[gray]{1.}$0.$& \cellcolor[gray]{
0.875}$12.5$& \cellcolor[gray]{0.75}$25.$& \cellcolor[gray]{
0.875}$12.5$& \cellcolor[gray]{0.5}$50.$\\\hline
\end{tabular}
\caption{Confusion matrix (in percentages) for the $z$-$y$-plane. The rows represent the sonified targets, the columns the selected field. The values indicate how often each field has been clicked when either of the targets was sonified.
}
\label{tab:confusion3}
\end{table*}

Tables \ref{tab:confusion1} to \ref{tab:confusion3} are tables of confusion. They show the relationship between sonified and clicked target fields not with a focus on the individual participant but on the individual field. Here, each row represents a sonified target field and the columns indicate the target as selected by the participants. The elements in the table indicate how frequently each field was marked as the target field by the participants. Consequently, the total of each row is $100$ \%. In addition to the numbers, the frequency is also indicated by gray level from white ($0$ \%) to black ($100$\%). The four quadrants are separated by double lines to highlight how many false clicks fall into the right quadrant. 
These tables help to get a further impression of whether the distributions of clicked targets differs significantly from random clicks, whether the participants performed similarly well in all three dimension pairs, and whether the distribution is roughly uniform across the whole two-dimensional spaces.

One can clearly see the dark diagonal lines, which indicates that the target field was typically correctly identified most frequently in all three groups. The correct field was identified in $25$ to $90.9$\% of all trials. Only in two to three out of $16$ cases per group the most frequently chosen field did not coincide with the sonified target field. Most confusions were between the target and other fields from the same quadrant. For most sonified targets, only three to four fields have been clicked at all. Only for very few targets, more than four different fields have been clicked by participants, namely t10 and t15 in group $x$-$y$, none in group $x$-$y$, and  t9 and t14 in group $z$-$y$. Some targets were only confused with one other target. This was the case for t1, t5, t11 in group $x$-$z$ and for t6, t7, t10 in group $z$-$y$. The $x$-$y$ group identified the outermost fields well, i.e., t4, t8, t12 and t16, and did not click on them, when any other target was sonified. The same is true for fields t4, t6, t9 and t13 for the $x$-$z$ group and for t3, t4 and t9 in the $z$-$y$ group. From visual inspection, all three tables seem similar to each other. This observation is confirmed by Kendall's $\tau$ test. After vectorization of the confusion matrices to one-dimensional vectors, the three show a fair but highly significant rank correlation ($\tau=0.56$, $p=4\times10^{-23}$ between $x$-$y$ and $x$-$z$, $\tau=0.49$, $p=2\times10^{-18}$ between $x$-$y$ and $z$-$y$, and $\tau=0.54$, $p=6\times10^{-21}$). This observation supports the finding from the ANOVA, i.e., that group did not have a significant effect on performance. The fair correlation is owed to the fact that all three share the strong diagonal. But the participants did not confuse the same fields in all three groups. 
The two groups with the new $z$ dimension exhibit similarities with the $x$-$y$ pair, which has already been shown to be orthogonal. This suggests that each dimension can be interpreted correctly during the presence of a second dimension, which indicates that all three dimensions are orthogonal.

All observations from the box plots \ref{pic:boxhit} to \ref{pic:boxz} and the confusion matrices \ref{tab:confusion1} to \ref{tab:confusion3} draw a coherent picture: We successfully identified and implemented three orthogonal auditory dimensions that are interpretable by inexperienced, passive listeners.

\section{Conclusion}
In this work we have highlighted the need of perceptually orthogonal dimensions for multidimensional or multivariate data sonification. We suggest five psychoacoustic quantities that can serve as three orthogonal dimensions. Experimental results show that these dimensions are learnable by inexperienced listeners under passive conditions. Participants were able to interpret the direction and distance of a sonified location in all two-dimensional pairs of the three-dimensional space. As our previous experiment on two-dimensional sonification \cite{jmui} has shown, the accuracy was much higher when participants interacted with the sound instead of listening passively. Furthermore, its resolution had been proven to be very high and the axes had been perceived as linear. Interactive experiments with the newly developed three-dimensional sonification will reveal whether the third dimension has the same qualities concerning accuracy, resolution and linearity.


\section{Outlook}
As ``(\ldots) is  critical  to  examine performance longitudinally when evaluating auditory display designs'' \cite{beacons}, we are designing a game to motivate users  for long-term interaction with the sonification. Progress on the game can be found on \url{http://curat.informatik.uni-bremen.de/}. In addition to interactive experiments to evaluate the sonification itself, we plan to evaluate the benefit of the sonification in a potential application area, like an image-guided surgery scenario.

\section{Key Points}
\begin{itemize}
\item We discussed the problem of low interpretability due to a lack of orthogonality in multidimentional/multivariate sonification
\item We identified a number of auditory attributes that seem perceptually orthogonal
\item We implemented them in a psychoacoustic sonification
\item Our experiment revealed that all three dimensions are in fact orthogonal to each other
\end{itemize}






\bibliographystyle{SageH}
\bibliography{hf2}

\end{document}